\documentclass[sigconf]{acmart}
\acmConference[EASE 2026]{The 30th International Conference on Evaluation and Assessment in Software Engineering}{9–12 June, 2026}{Glasgow, Scotland, United Kingdom}

\usepackage{balance}

\AtBeginDocument{%
  }

\usepackage{algorithmic}
\usepackage{graphicx}
\usepackage{textcomp}
\usepackage{xcolor}
\usepackage{flushend}
\usepackage{url}
\usepackage{svg} 
\usepackage{subcaption} 

\usepackage{booktabs}
\usepackage{multirow}
\usepackage{framed}
\usepackage{hyperref}
\usepackage{array}
\usepackage{siunitx} 
\usepackage{tabularx}
\usepackage{makecell} 
\usepackage{caption}
\usepackage{enumitem}

\usepackage{nimbusmononarrow}
\usepackage[T1]{fontenc}

\usepackage{tikz}
\newcommand*\circled[1]{\tikz[baseline=(char.base)]{
            \node[shape=circle,draw,inner sep=1pt] (char) {#1};}}

\usepackage{afterpage}

\newcolumntype{C}[1]{>{\centering\arraybackslash}p{#1}}

\newcommand{\repo}[1]{\texttt{\small #1}}

\usepackage{seqsplit}
\renewcommand{\texttt}[1]{{\ttfamily\seqsplit{#1}}}

\setcopyright{acmlicensed}
\copyrightyear{2018}
\acmYear{2018}
\acmDOI{XXXXXXX.XXXXXXX}
\acmISBN{978-1-4503-XXXX-X/2018/06}

\begin{document}

\title{Beyond the YAML File: Understanding Real-World GitHub Actions Workflow Adoption}

\author{Ali Khatami}
\email{s.khatami@tudelft.nl}
\orcid{0000-0002-2212-2311}
\affiliation{%
  \institution{Delft University of Technology}
  \city{Delft}
  \country{The Netherlands}
}
\author{Carolin Brandt}
\email{c.e.brandt@tudelft.nl}
\orcid{0000-0001-7623-1970}
\affiliation{%
  \institution{Delft University of Technology}
  \city{Delft}
  \country{The Netherlands}
}
\author{Andy Zaidman}
\email{a.e.zaidman@tudelft.nl}
\orcid{0000-0003-2413-3935}
\affiliation{%
  \institution{Delft University of Technology}
  \city{Delft}
  \country{The Netherlands}
}





\renewcommand{\shortauthors}{A. Khatami, C. Brandt, A. Zaidman}

\begin{abstract}
Continuous Integration and Continuous Deployment (CI/CD) have become fundamental to modern software development, with GitHub Actions (GHA) emerging as a dominant automation platform.
In this study, we analyze real-world execution records of GHA, examining how developers react to workflow failures, how these workflows are utilized by projects, and how these aspects relate to project characteristics.
We quantitatively analyze 258,300 workflow run records from 952 repositories and perform an in-depth qualitative analysis of 21 selected, diverse GitHub repositories to understand how  maintainers and contributors interact with workflow results.
We identify three distinct failure response patterns, observe that higher usage intensity of GHA workflows correlates with lower failure rates, and uncover a configuration-usage gap where the presence of configuration files masks disabled or unused workflows.
Moreover, our qualitative analysis of relationships between project characteristics and utilization patterns yields five hypotheses for future validation.
\end{abstract}




\maketitle

\section{Introduction}
Continuous Integration and Continuous Deployment (CI/CD) have become 
pillars of modern software development, enabling teams to automate testing, building, 
and deployment processes~\cite{bellerMSR2017,bellerMSR2017a,HiltonASE2016,elazharyICSME2019,elazharyTSE2022,Fowler2006,klotinsEMSE2022,santosEMSE2025}. GitHub Actions (GHA), introduced in 2019, has rapidly emerged as one 
of the most widely adopted CI/CD platforms, offering developers a powerful yet 
accessible framework for workflow automation directly integrated with their source code 
repositories~\cite{DBLP:conf/icsm/DecanMMG22,zhangICSE2024,khatamiSCAM2024b}. While the potential benefits of workflow 
automation are well-documented~\cite{HiltonASE2016,vasilescuESECFSE2015}, realization of these benefits depends on how development teams utilize tools in practice~\cite{khatami2023state}.

Existing studies~\cite{DBLP:conf/icse/BouzeniaP24, DBLP:conf/icsm/DecanMMG22, DBLP:journals/ese/WesselVGT23} analyze static workflow configuration (YAML) files defining automation steps, taking the presence of configuration files as a proxy for usage.
However, studies on static analysis tools revealed a significant gap between configured tools and their actual usage~\cite{bholanathSANER2016,johnsonICSE2013}.
Many projects maintain configuration files for tools that are not or rarely used, or whose results are consistently ignored by developers~\cite{besseyCACM2010}.
This disconnect suggests that configuration files alone provide an incomplete view of how automation technologies impact software development processes.

In this study, we instead examine \textit{workflow run data}: the execution records of GHA workflows.
This includes the events that trigger runs and run outcomes. We qualitatively analyze the responses of developers to failures through subsequent commits and pull request (PR) discussions to develop a richer understanding of GHA's role in modern software development processes.
We address the following research questions:

\begin{description}
    \item[\textbf{RQ1}] \textbf{What are the observed\\GitHub Actions utilization patterns?} 
    \end{description}
    Following a quantitative approach for RQ1, we examine utilization intensity (number of runs and failure rates) and trigger patterns (events causing workflow executions). This reveals how frequently failures occur and whether intensive workflow adoption correlates with failure rates. 

    \begin{description}
    \item[\textbf{RQ2}] \textbf{How do developers react to\\GitHub Actions workflow failures?} 
    \end{description}
    To understand the effect of GHA adoption on development processes, we first focus on failure points in pull requests and main branches, as they enable us to analyze subsequent developer actions that address these failures.

    \begin{description}
    \item[\textbf{RQ3}] \textbf{What relationships exist between project characteristics and GitHub Actions utilization patterns?} 
\end{description}
    \noindent
    For RQ3 we examine relationships between project characteristics and the patterns identified in RQ2 and RQ1 through a qualitative analysis of 21 repositories. Given the complexity beyond simple categorical mappings, we propose hypotheses for future validation rather than definitive conclusions.

To answer these questions, we 
quantitatively analyze run records from 765 repositories based on a dataset of 952 GitHub projects~\cite{DBLP:conf/icse/BouzeniaP24}.
Then we systematically select 21 repositories for a qualitative analysis
based on three key dimensions: popularity, workflow run frequency, and failure rates.
We conduct in-depth analysis of each repository, examining workflow run records, developer interactions in PRs, associations between workflow runs and commits (both in PRs and the main branch), and the contextual characteristics of projects. 
Previous studies established that quality assurance (QA)
practices vary considerably across different development contexts~\cite{khatamiSCAM2024}. The adoption and effectiveness of these practices are influenced by numerous factors including team size, technology stack, and development methodology. GHA workflows, as a mechanism for automating QA activities, likely exhibit similar contextual dependencies.

By examining how development teams interact with GitHub Actions in practice, our research \textbf{contributes a more nuanced understanding of GitHub Actions workflow adoption} than configuration analysis alone could provide. 
We identify patterns in how developers respond to workflow failures, recognizing common workflow utilization patterns, and examine relationships between these patterns and project characteristics. 
These findings can be the basis for future studies of GHA practices, grounding them in rich observational data.

Our paper is structured as follows: in Section~\ref{section:background} we review  GitHub Actions studies. We then describe our methodology for our mixed-methods analysis in Section~\ref{section:methodology}. 
In Sections~\ref{section:RQ1} through~\ref{section:RQ3} we present our findings regarding workflow utilization patterns across different project contexts (RQ1), developer responses to workflow failures (RQ2), and we explore relationships between these patterns and project characteristics (RQ3). Section~\ref{section:discussion} presents a discussion of the implications of our study.
We conclude our paper with Section~\ref{section:conclusion}.

\section{Background}\label{section:background}
GitHub Actions (GHA), introduced in 2019, has rapidly become the dominant Continuous Integration and Continuous Delivery (CI/CD) automation tool within the GitHub ecosystem, largely due to its deep integration and the provision of free resources to open source projects~\cite{delicheh2023preliminary, golzadeh2022rise, rostami2023usage}. 
Quantitative studies of GHA configuration files demonstrated widespread and growing adoption: 22\% of popular projects adopted GHA according to early studies~\cite{chen2021let}, increasing to 37\% for popular GitHub projects and up to 57\% for repositories utilizing the most popular programming languages~\cite{ayala2023empirical}.
Previous research explored various facets of GHA adoption:

\subsection{Impact on Development Practices} 
Research on the impact of GHA shows that its adoption generally correlates with faster resolution latency for pull requests (PRs) and issues, and increased commit frequency~\cite{chen2021let}. 
However, other studies report a more nuanced picture, noting an increase in rejected PRs and fewer commits in merged PRs~\cite{DBLP:journals/ese/WesselVGT23,kinsman2021software}, with accepted PRs receiving more discussion comments and taking longer to merge after adoption~\cite{DBLP:journals/ese/WesselVGT23}.

\subsection{Ecosystem Structure and Evolution}
Studies describe GHA workflows and reusable components (Actions) as a complex, rapidly evolving ecosystem~\cite{DBLP:conf/icsm/DecanMMG22, decan2023outdatedness}. 
Workflows, defined by jobs and steps, undergo continuous modification~\cite{DBLP:conf/icsm/DecanMMG22} including fixing YAML syntax errors, debugging, and updating instructions, leading researchers develop specialized tools for tracking commit changes of  workflow configurations~\cite{rostami2024gawd, valenzuela2022evolution}.
Although action reuse is widespread~\cite{DBLP:conf/icsm/DecanMMG22, decan2023outdatedness}, most workflows reference outdated releases, typically lagging at least seven months behind the latest version~\cite{decan2023outdatedness}.
Moreover, analyses estimate an average annual cost of \$504 for GHA adoption of paid-tier repositories, while optimizations in resource consumption are underutilized~\cite{DBLP:conf/icse/BouzeniaP24}. 

\subsection{Security and Vulnerability Management}
A major focus of studies has been the security risks in the GHA workflow configurations~\cite{ayala2023empirical, delicheh2023preliminary,benedetti2022automatic,delicheh2024mitigating,koishybayev2022characterizing}.  
99.8\% of workflows are over-privileged with default read-write access, and 23.7\% are exploitable for arbitrary code execution via PRs~\cite{koishybayev2022characterizing}. 
Vulnerable third-party actions are common~\cite{koishybayev2022characterizing}, and security recommendations such as using commit hashes are followed by fewer than 3\% of repositories~\cite{decan2023outdatedness}.  
Security tools like CodeQL remain underused; only 13.5\% of top repositories enabling CodeQL actually use it~\cite{ayala2023empirical}. Developers also cite security and complexity concerns when selecting actions~\cite{saroar2023developers}.  

Most empirical research statically analyzes workflow configuration (YAML) files  found in the \textit{.github/workflows} directory, using their presence as a proxy for adoption and usage~\cite{DBLP:conf/icse/BouzeniaP24, DBLP:conf/icsm/DecanMMG22, DBLP:journals/ese/WesselVGT23}.
Only Bouzenia and Pradel~\cite{DBLP:conf/icse/BouzeniaP24} focus on resource optimization of GHA workflows.
The novelty of this paper is to analyze workflow run data, i.e., the execution records of GHA workflow runs including their trigger events and outcomes with qualitative analysis of developers' reactions to them. This approach is necessary to understand actual workflow adoption beyond configuration presence.

\begin{figure*}[htbp]
\centering
\includegraphics[width=1\textwidth]{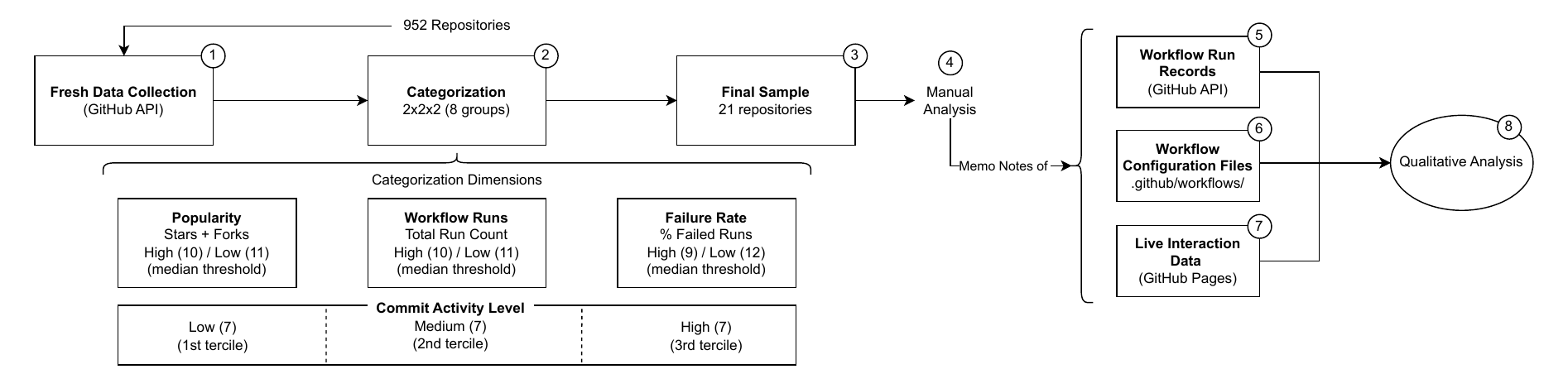}
\vspace{-6.5mm}
\caption{Methodology Overview} 
\label{fig:methodology-overview}
\vspace{-3mm}
\end{figure*}

\section{Methodology}\label{section:methodology}
We conducted an exploratory analysis of GitHub Actions (GHA) workflow adoption through a mixed methods analysis of workflow runs. This section explains our data collection process, repository sampling strategy, and the manual analysis approach used to answer our research questions.
An overview 
is depicted in Figure~\ref{fig:methodology-overview}.

Our study focuses on analyzing execution data from GHA workflows and related contextual information of repositories. We examine the recorded workflow runs\footnote{``Workflow Run'' is the exact terminology used by GitHub, more details of these records are available at \url{https://docs.github.com/en/rest/actions/workflow-runs?apiVersion=2022-11-28\#get-a-workflow-run}} from the GitHub API, whether these runs pass or fail, how they were triggered, their associated pull requests (PRs) or commits, and how developers react to these outcomes. 
We define \textit{workflow run data} as the execution history of GHA workflows, including timestamps, trigger events\footnote{Events that trigger workflows: \url{https://docs.github.com/en/actions/reference/workflows-and-actions/events-that-trigger-workflows\#about-events-that-trigger-workflows}}, GitHub page URLs, status and conclusion, associated commits, PRs, users, and branches. 
We examine these execution records in combination with observable developer reactions on PR and branch web pages. 

\subsection{Data Collection}
Our goal is to identify GitHub repositories that actively use GHA and, thus, have workflow run data available for analysis. This presents two challenges: (1) finding a representative sample of repositories using GHA, and (2) ensuring these repositories have accessible execution history. To address these challenges, we begin with an existing dataset while recognizing that GitHub's workflow run retention policy\footnote{\url{https://docs.github.com/en/actions/administering-github-actions/usage-limits-billing-and-administration\#workflow-run-history-retention-policy}} would require us to collect fresh data.

We use the dataset by Bouzenia and Pradel~\cite{DBLP:conf/icse/BouzeniaP24}, which contains 952 GitHub repositories with workflow run histories. This dataset has a balanced representation of popular repositories (600 with 100+ stars and commits) and less popular ones (352 with fewer than 100 stars). 
Of the initial 952 repositories, 946 were still publicly available 
at the time of data collection. We used the GitHub API to collect fresh workflow run data from these 946 repositories, of which 765 contained workflow runs. 
To collect sufficient data while staying within reasonable API usage limits, we collected up to 1000 run records per repository\footnote{This threshold reflects both API constraints and our pattern-identification goal. Collecting beyond 1000 runs per repository would require more resource and time while providing limited additional insight, as high-volume repositories represent a small minority with predominantly bot-triggered workflows.}, yielding 258,300 run records collected at the end of 2024 (Figure~\ref{fig:methodology-overview} -- \circled{1}). 
We filtered 213,640 non-bot-triggered runs across 742 repositories to analyze for RQ1~(Section~\ref{section:RQ1}).

Our data collection pipelines, available in our replication package~\cite{khatamiSANERReplicationPackage}, queried the GitHub API for recent workflow runs, commit histories, and other repository metadata, saving them into a MongoDB database~\cite{khatamiSANERReplicationPackage} for subsequent analysis. 

\begin{table*}[!t]
\centering
\footnotesize
\setlength{\tabcolsep}{4pt}
\caption{Sampled Repositories Categorization Attributes}
\vspace{-2mm}
\begin{tabular}{@{}clcrrrcrcc@{}}
\toprule
\textbf{ID} & \textbf{Repository} & \textbf{Pop.} & \textbf{Pop.G} & \textbf{Comm.} & \textbf{Comm.G} & \textbf{Runs} & \textbf{Run.G} & \textbf{Run.Fail\%} & \textbf{Run.Fail.G} \\
\midrule
R1 & trusttoken/contracts-pre22 & 450 & low & 18 & low & 46 & low & 0.0 & low \\
R2 & nextcloud/ocsms & 240 & low & 32 & medium & 7 & low & 0.0 & low \\
R3 & ngageoint/geopackage-js & 393 & low & 37 & medium & 21 & low & 14.3 & low \\
R4 & yandex-cloud/serverless-plugin & 72 & low & 91 & medium & 37 & low & 8.1 & low \\
R5 & iouAkira/someDockerfile & 181 & low & 15 & low & 15 & low & 86.7 & high \\
R6 & dotMorten/NmeaParser & 351 & low & 16 & low & 48 & low & 47.9 & high \\
R7 & nemuTUI/nemu & 370 & low & 20 & low & 126 & low & 6.3 & low \\
R8 & data-driven-forms/react-forms & 399 & low & 191 & high & 64 & low & 4.7 & low \\
R9 & komamitsu/fluency & 190 & low & 75 & medium & 234 & high & 16.2 & high \\
R10 & aws-cloudformation/cfn-lint-vscode & 416 & low & 103 & medium & 1000 & high & 2.7 & low \\
R11 & boutproject/BOUT-dev & 282 & low & 644 & high & 1000 & high & 20.5 & high \\
R12 & felix-fly/v2ray-openwrt & 823 & high & 7 & low & 2 & low & 0.0 & low \\
R13 & Thinkmill/manypkg & 974 & high & 47 & medium & 142 & high & 12.7 & low \\
R14 & webextension-toolbox/webext-toolbox & 800 & high & 27 & low & 735 & high & 21.2 & high \\
R15 & prontolabs/pronto & 2874 & high & 14 & low & 29 & low & 0.0 & low \\
R16 & ai/size-limit & 8403 & high & 160 & high & 147 & high & 41.5 & high \\
R17 & lc-soft/LCUI & 4525 & high & 209 & high & 135 & low & 74.8 & high \\
R18 & tsl0922/ttyd & 9214 & high & 73 & medium & 365 & high & 0.5 & low \\
R19 & RailsEventStore/rails\_event\_store & 1546 & high & 1151 & high & 1000 & high & 0.2 & low \\
R20 & BeyondDimension/SteamTools & 22097 & high & 1466 & high & 1000 & high & 25.6 & high \\
R21 & libcpr/cpr & 7599 & high & 294 & high & 1000 & high & 30.4 & high \\
\bottomrule
\end{tabular}
\label{tab:repo-categories}
\\
Pop: Popularity (stars + forks), Pop.G: Popularity group, Comm: Total commits since 2023, Comm.G: Total commits group\\
Run.G: Run group, Run.Fail\%: Failure rate percentage, Run.Fail.G: Failure group
\vspace{-3mm}
\end{table*}

\begin{figure*}[t]
\centering
\includegraphics[width=0.24\textwidth]{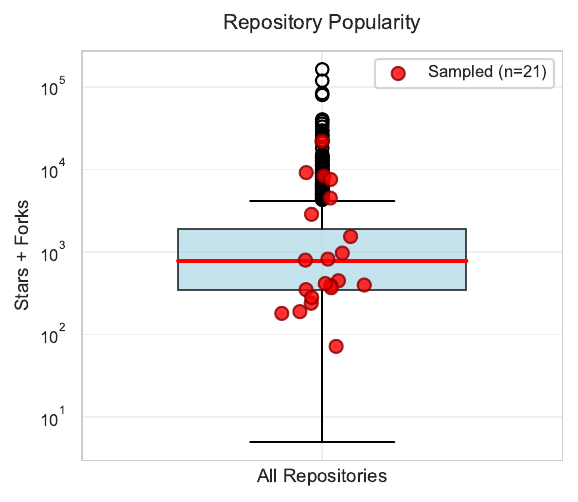}\hfill
\includegraphics[width=0.24\textwidth]{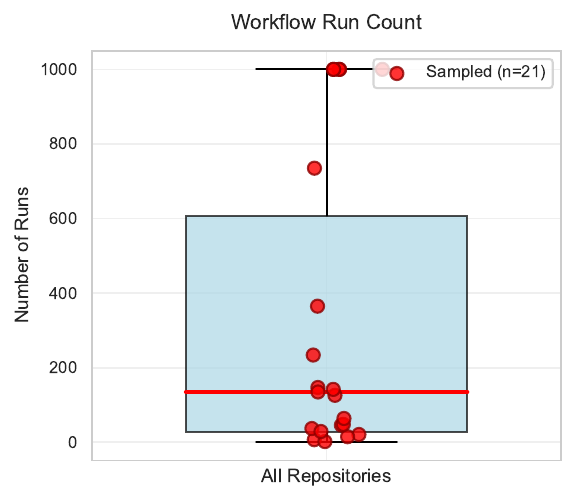}\hfill
\includegraphics[width=0.24\textwidth]{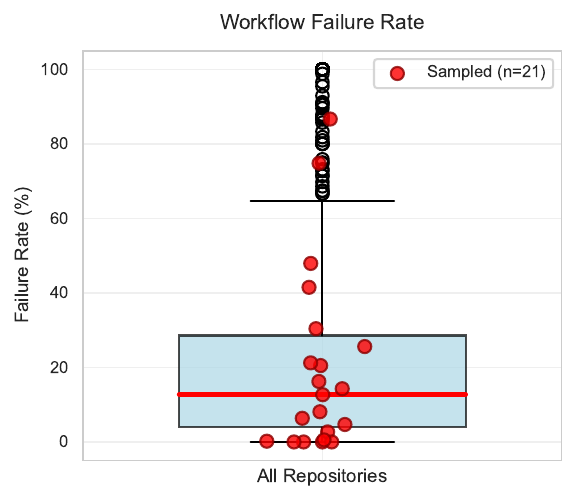}\hfill
\includegraphics[width=0.24\textwidth]{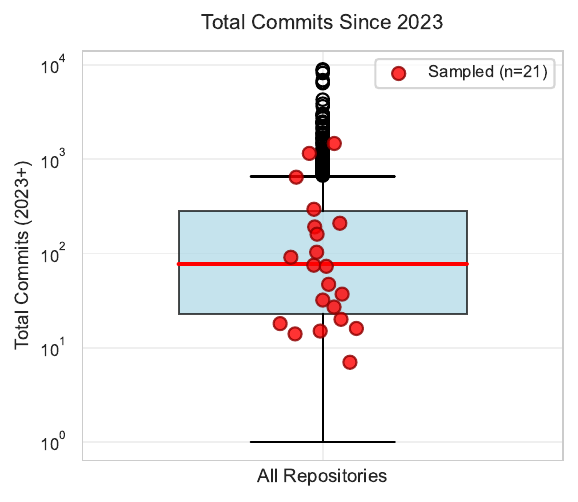}
\vspace{-3mm}
\caption{Distribution of sampled repositories (red dots, n=21) across the population. Our sample spans the complete range across (a) repository popularity, (b) workflow run count, (c) failure rates, and (d) commit activity since 2023.}
\label{fig:four-figures}
\vspace{-5mm}
\end{figure*}

\subsection{Categorization and Sampling Strategy}
\label{sec:categorization-and-sampling}
To capture 
GHA adoption patterns, we classified the 765 repositories with workflow runs using three attributes: popularity (stars + forks), workflow activity (total run count), and workflow failure rate (percentage of failed runs). We chose these dimensions to represent 
project scales, CI/CD adoption levels, and failure response contexts essential for our research questions (Figure~\ref{fig:methodology-overview} -- \circled{2}).

\begin{table}[!b]
\centering
\vspace{-3mm}
\caption{Repository Sampling: 2×2×2 Classification}
\vspace{-2mm}
\label{tab:sampling_classification}
\footnotesize
\begin{tabular}{ccccc}
\toprule
\textbf{Group} & \textbf{Pop} & \textbf{\%} & \textbf{Sampled Repositories} & \textbf{n} \\
\midrule
L-L-L & 105 & 13.7 & R1, R2, R3, R4, R7 & 5 \\
L-L-H & 112 & 14.6 & R5, R6 & 2 \\
L-H-L & 86 & 11.2 & R8, R10 & 2 \\
L-H-H & 80 & 10.5 & R9, R11 & 2 \\
H-L-L & 72 & 9.4 & R12, R15 & 2 \\
H-L-H & 94 & 12.3 & R17 & 1 \\
H-H-L & 120 & 15.7 & R13, R18, R19 & 3 \\
H-H-H & 96 & 12.6 & R14, R16, R20, R21 & 4 \\
\midrule
\textbf{Total} & \textbf{765} & \textbf{100} & & \textbf{21} \\
\bottomrule
\end{tabular}
\footnotesize
\\Pop=Population count; \%=Population percentage\\
Groups: Popularity-Runs-FailureRate (L=Low, H=High)
\end{table}

Each attribute was split into high/low categories using median thresholds, creating 8 combinations (2×2×2) as shown in Table~\ref{tab:sampling_classification}. Analysis of the population showed relatively balanced distribution across these groups, ranging from 9.4\% to 15.7\% of repositories per group, outlining that no single usage pattern dominates. While one repository per group could provide insight into each GHA adoption pattern, we selected multiple repositories from most groups based on observed diversity within groups and the need to capture varied adoption approaches within each categorization groups.

\vspace{1mm}
\noindent
\textbf{Sampling Approach:}
we selected 21 repositories using purposive sampling to ensure representation across all 8 classification groups (Table~\ref{tab:sampling_classification} and Figure~\ref{fig:methodology-overview} -- \circled{3}). 
Rather than proportional sampling, we varied the number of repositories per group (1--5 repositories) based on research value, where some patterns (e.g., low-activity projects with low number of runs and high-activity with high number of runs) needed more examples to understand diverse utilization approaches, and practical coverage to ensure every combination was represented without missing important behavioral patterns.

Figure~\ref{fig:four-figures} demonstrates that our sampled repositories span the distribution across all three classification dimensions (popularity, run counts, and run failure rates), ensuring coverage of diverse values. While commit activity was not part of our classification, our sample spans development activity levels capturing projects with varying development intensities.

\vspace{1mm}
\noindent
\textbf{Sample Characteristics:}
our final sample ranges from less popular/active projects (72 stars+forks, 7 commits) to more popular/active ones (22,097 stars+forks, 1,466 commits), with 2-1,000 workflow runs and failure rates from 0\% to 86.7\% (Table~\ref{tab:repo-categories}). This diversity lets us analyze how GHA adoption and developer behaviors vary across different project contexts.
We intentionally keep repositories with very few runs, as we are also interested to study project that have workflows configured but rarely use them.

We enriched this data with contextual attributes, including projects' programming languages, commit count, workflow configuration commit count, team and project size indicators to facilitate deeper understanding of project contexts during our sampling and manual analysis phases (Figure~\ref{fig:methodology-overview} -- \circled{3}). 
These additional attributes are detailed in Table~\ref{tab:repo-additional}. The complete dataset is available in~\cite{khatamiSANERReplicationPackage}. 

\subsection{Manual Analysis through Qualitative Coding}\label{sec:manual-analysis}

To answer RQ2 and RQ3, we conduct a manual analysis.
For each of the 21 selected repositories, we analyzed workflow run records (Figure~\ref{fig:methodology-overview} -- \circled{5}), workflow configuration files (Figure~\ref{fig:methodology-overview} -- \circled{6}), and live interaction data\footnote{Any form of interactions observable from the GitHub website UI, e.g., comments in pull requests, commit messages, active/deactivated workflows.} (Figure~\ref{fig:methodology-overview}~--~\circled{7}), 
both from our API-collected data (see our dataset~\cite{khatamiSANERReplicationPackage}) and live data on the repositories' GitHub page.
Our manual analysis of each repository includes:
\begin{enumerate}
    \item \textbf{Workflow file configuration:} 
    analyzing purpose of configuration files (\textit{.github/workflows/*.yml}) and their commit history if and why they were updated.
    \item \textbf{Workflow run records:} examining workflow run outcomes on main branches and in pull requests, with particular attention to 
    developers' reactions to  failures.
    \item \textbf{Developer interaction:} analyzing commit messages after failures and discussions in PRs to understand if and how workflow failures triggered actions.
    \item \textbf{Repository contextual information:} analyzing characteristics like programming language, team size, maintainer count, development style~\cite{gousios2014pullbased}.
\end{enumerate}

Our qualitative analysis (Figure~\ref{fig:methodology-overview} -- \circled{8}) starts with open coding and memo-writing by the first two authors for each repository from the mentioned elements above. 
Both authors analyzed repositories separately before merging the codes and grouping them into higher-level codes.
These were then categorized into broader themes through constant comparison~\cite{glaser2017discovery} with the original memos. The authors discussed the emerging groups until reaching a negotiated agreement~\cite{DBLP:journals/iahe/GarrisonCKK06}. 
The analysis required substantial manual effort to scavenge data from multiple sources and interpret the collected information through systematic coding.

\subsection{Answering the research questions}
Our systematic qualitative coding process of 21 repositories, together with the quantitative analysis of 765 repositories with run records, enabled us to understand workflow adoption beyond their configuration files by examining the rich contextual data surrounding workflow adoption and developer interactions. 
We use this mixed-methods analysis to answer our three research questions in Sections~\ref{section:RQ1} to~\ref{section:RQ3}.

\begin{table*}
\centering
\footnotesize
\setlength{\tabcolsep}{3pt}
\caption{Repository Additional Information}
\vspace{-2mm}
\begin{tabular}{@{}clcclcllrr@{}}
\toprule
\textbf{ID} & \textbf{Repository} & \textbf{Q.Avg} & \textbf{WF.C} & \textbf{Lang} & \textbf{Age} & \textbf{Trigger Event Distribution} & \textbf{Mo.Gap} & \textbf{U.Runs} & \textbf{B.Runs} \\
\midrule
R1 & trusttoken/contracts-pre22 & 0.0 & 0 & TypeScript & 7.1 & sch: 38, pr: 8 & 8.3 & 46 & 0 \\
R2 & nextcloud/ocsms & 0.0 & 0 & JavaScript & 10.3 & sch: 7 & 1.4 & 7 & 0 \\
R3 & ngageoint/geopackage-js & 0.5 & 4 & TypeScript & 9.1 & push: 19, dyn: 2 & 2.6 & 19 & 2 \\
R4 & yandex-cloud/serverless-plugin & 0.0 & 0 & TypeScript & 4.7 & pr: 18, push: 15, dyn: 4 & 14.2 & 20 & 17 \\
R5 & iouAkira/someDockerfile & 0.7 & 6 & Python & 5.4 & push: 12, wd: 3 & 9.3 & 15 & 0 \\
R6 & dotMorten/NmeaParser & 1.2 & 10 & C\# & 10.9 & push: 24, pr: 16, wd: 5, dyn: 3 & 13.0 & 48 & 0 \\
R7 & nemuTUI/nemu & 0.2 & 2 & C & 5.2 & push: 72, pr: 54 & 13.1 & 126 & 0 \\
R8 & data-driven-forms/react-forms & 0.0 & 0 & JavaScript & 5.8 & pr: 62, dyn: 2 & 15.1 & 51 & 13 \\
R9 & komamitsu/fluency & 0.5 & 4 & Java & 9.3 & pr: 196, push: 32, dyn: 6 & 13.0 & 50 & 184 \\
R10 & aws-cloudformation/cfn-lint-vscode & 2.0 & 17 & JavaScript & 6.7 & sch: 878, pr: 78, push: 40, rel: 4 & 9.6 & 945 & 55 \\
R11 & boutproject/BOUT-dev & 8.7 & 72 & C++ & 11.4 & push: 489, pr: 477, dyn: 20, sch: 14 & 3.3 & 880 & 120 \\
R12 & felix-fly/v2ray-openwrt & 0.5 & 4 & Shell & 6.3 & push: 2 & 6.9 & 2 & 0 \\
R13 & Thinkmill/manypkg & 0.5 & 4 & TypeScript & 5.4 & pr: 82, push: 60 & 11.7 & 136 & 6 \\
R14 & webextension-toolbox/webext-toolbox & 0.7 & 6 & TypeScript & 7.0 & pr: 572, sch: 67, push: 49, dyn: 46, wd: 1 & 15.3 & 159 & 576 \\
R15 & prontolabs/pronto & 0.5 & 4 & Ruby & 11.6 & pr: 22, push: 7 & 14.8 & 29 & 0 \\
R16 & ai/size-limit & 1.8 & 15 & JavaScript & 7.6 & push: 90, pr: 57 & 12.8 & 144 & 3 \\
R17 & lc-soft/LCUI & 3.2 & 27 & C & 12.5 & push: 134, pr: 1 & 13.9 & 133 & 2 \\
R18 & tsl0922/ttyd & 1.0 & 8 & C & 8.4 & push: 220, pr: 145 & 13.8 & 135 & 230 \\
R19 & RailsEventStore/rails\_event\_store & 13.0 & 108 & Ruby & 9.8 & sch: 960, push: 28, pr: 12 & 1.9 & 998 & 2 \\
R20 & BeyondDimension/SteamTools & 3.0 & 25 & C\# & 4.1 & push: 791, pr: 98, create: 69, delete: 42 & 13.1 & 901 & 99 \\
R21 & libcpr/cpr & 4.8 & 40 & C++ & 9.8 & pr: 454, push: 426, dyn: 117, rel: 2, wd: 1 & 7.1 & 883 & 117 \\
\bottomrule
\end{tabular}
\label{tab:repo-additional}
\\Q.Avg: Quarterly avg workflow commits, WF.C: Workflow commits since 2023, Mo.Gap: Months between first and last run record\\
U/B.Runs: User/Bot runs. Event types: pr: pull\_request, sch: schedule, dyn: dynamic, wd: workflow\_dispatch, rel: release
\vspace{-3mm}
\end{table*}

\section{\textbf{RQ1: What are the observed GitHub Actions utilization patterns?}}
\label{section:RQ1}
Beginning our analysis, we investigate the broader patterns of how teams utilize GitHub workflows across different repositories.
We conduct quantitative analysis across 765 repositories with existing run records, examining the \textbf{utilization intensity spectrum} (run counts and failure rates) and \textbf{workflow trigger patterns} (events triggering workflow runs). This reveals how frequently run failures occur, whether intensive workflow adoption correlates with failure rates, and which event types and combinations trigger workflow executions in practice.

The workflow trigger analysis 
shows us what triggers runs based on 
the run records, rather than their configuration.
In this way, we can distinguish whether workflows are activated as part of maintenance automation, or through development-driven triggers like a push or a pull request.

\subsection{Utilization Intensity Spectrum}\label{utilization-intensity-spectrum}

\begin{figure}[!t]
\centering
\includegraphics[width=0.45\textwidth]{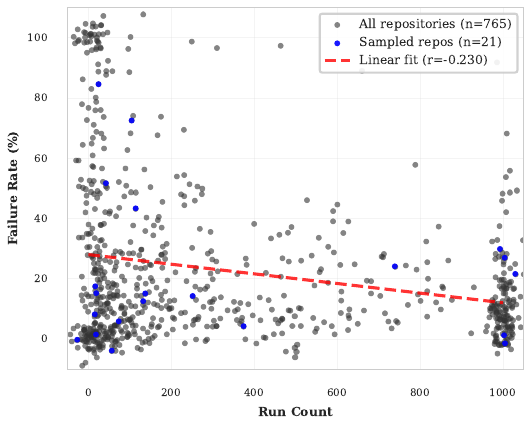}
\vspace{-2.5mm}
\caption{Repositories Run Count vs. Run Failure Rate. Points are jittered horizontally to reduce overplotting. 
}
\vspace{-4.5mm}
\label{fig:run-count-vs-failure}
\end{figure}

Repositories demonstrate an intensity spectrum from minimal to extensive workflow usage. Pearson correlation~\cite{dekking2005modern} analysis reveals a statistically significant negative correlation between run count and failure rate (Correlation coefficient $r 
= -0.230$, $p < 0.001$ across 765 repositories): as workflow usage increases, failure rates tend to decrease. 
This high usage can result from two distinct patterns: frequent triggering of workflows due to active development (\emph{activity-intensive}), or comprehensive workflow configurations that execute many jobs per trigger (\emph{configuration-intensive}). In both cases, whether 50 workflow runs result from one event triggering 50 jobs or 50 separate events each triggering one job, the total run volume contributes to the observed statistically significant negative correlation in Figure~\ref{fig:run-count-vs-failure}.

To discuss the different trends in Figure~\ref{fig:run-count-vs-failure}, we split the resitories into three visually distinct but roughly equally sized buckets:
low-usage (<100 runs, 45\%), medium-usage (100-499 runs, 26.5\%), and high-usage ($\geq$500 runs, 28.5\%).
These thresholds separate three observable behavioral patterns in Figure~\ref{fig:run-count-vs-failure}: 
the left cluster shows sporadic usage with high failure variability, 
the right cluster shows sustained automation with more stable failure rates, 
and the middle represents a transitional pattern\footnote{\textbf{Note on threshold selection:} The numerical thresholds (100 and 500 runs) presented in this section serve as descriptive categories for interpreting patterns within our specific sample, not as generalizable classification boundaries. These values emerged from visual analysis of our data distribution and are specific to our collection period and sampling approach. The primary contribution is identifying the pattern of systematically decreasing failure rates with increased usage intensity; the specific threshold values are artifacts of our sample and should not be applied to other contexts.}.

\noindent
\textbf{High-Usage Pattern} (28.5\% with $\geq$500 runs): 
these repositories show high run volumes through two distinct approaches. 
\texttt{RailsEventStore/rails\_event\_store} exemplifies the \emph{activity-intensive}
approach with 1000 runs and 0.2\% failure rate during two months, generating volume through frequent development activity supported by 960 scheduled runs\footnote{The schedule event allows to trigger a workflow at a scheduled time, like a cron job.} and an explicit culture of keeping all runs passing. \texttt{boutproject/BOUT-dev} represents the \emph{configuration-intensive} approach with 1000 runs and 20.5\% failure rate across 50+ workflow files during 3.3 months, reaching high run record volume through automation breadth while tolerating higher failure rates (e.g., ignoring platform-specific failures in PR \#3029, \textit{``LGTM, fedora failure is unrelated''}). 

\noindent
\textbf{Low-Usage Pattern} (45\% with $<$100 runs): these repositories show extreme variability in their run outcomes. Successful minimal usage includes \texttt{felix-fly/v2ray-openwrt} (2 runs, 0\% failure), \texttt{prontolabs/pronto} (29 runs, 0\% failure), and \texttt{trusttoken/contracts-pre22} (46 runs, 0\% failure), exemplifying focused minimal automation. 
High failure rate  
minimal usage includes \texttt{iouAkira/someDockerfile} (15 runs, 86.7\% failure rate): this project creates new workflow files per change and keeps pushing changes till they pass (using it as a tool to publish to docker hub).
In \texttt{dotMorten/NmeaParser} (48 runs, 31.3\% failure rate) the maintainer keeps pushing changes to fix failing workflows after making changes, which results in many workflow failure records. These usage patterns with high failure rates show experimental or trial-and-error approaches.

\noindent
\textbf{Medium-Usage Pattern} (26.5\% with 100-499 runs): these repositories show transitions between low and high-usage patterns. \texttt{webextension-toolbox/webextension-toolbox} shows 735 runs with a 21.2\% failure rate, where 149 of 156 failures were bot-triggered, the PRs were then closed by the bots (e.g., \#1071, \#1070, \#1068, \#1067, \#1065, etc.) indicating automation complexity without observable influence on the development. \texttt{libcpr/cpr} demonstrates 883 runs with 6.9\% failure rate while explicitly accepting failures (PR \#1170, \textit{``CI failures are expected and I will fix them soon''}). 

The key finding is that average failure rates decrease with higher usage intensity. 
This pattern suggests that sustained utilization of workflow automation (whether through frequent usage or comprehensive configuration) correlates with lower failure rates, while sporadic or experimental usage results in more variable failure rate.

\subsection{Workflow Trigger Patterns}\label{workflow-trigger-patterns}
Table~\ref{tab:trigger-event-details}-a summarizes our analysis of \num{213640} non-bot-triggered\footnote{Based on what we see in RQ2, we decided to exclude bot-triggered runs because they showed minimal meaningful reactions compared to user-triggered runs and could skew our results.} runs in 742 repositories, 
showing that teams primarily use development triggers, with selective adoption of scheduled triggers. 

\begin{table}[!t]
\centering
\caption{Trigger events: distribution and combinations}
\vspace{-2mm}
\label{tab:trigger-event-details}
\scriptsize
\begin{tabular}{@{}lr@{\hspace{0.3em}}r@{\hspace{1em}}lr@{\hspace{0.3em}}r@{}}
\toprule
\multicolumn{3}{c}{\textbf{(a) Top 10 Trigger Event Distribution}} & \multicolumn{3}{c}{\textbf{(b) Top 10 Trigger Event Comb.}} \\
\cmidrule(r){1-3} \cmidrule(l){4-6}
\textbf{Event Type} & \textbf{Runs} & \textbf{\%} & \textbf{Event Combination} & \textbf{Repos} & \textbf{\%} \\
\midrule
push                  & 76,822        & 36        & PR + push                  & 238            & 32                              \\
pull\_request         & 76,400        & 36        & push only                  & 85             & 11                              \\
schedule              & 39,637        & 19        & PR + push + sched          & 67             & 9                               \\
dynamic               & 6,387         & 3        & dyn + PR + push            & 40             & 5                               \\
pull\_request\_target & 4,399         & 2         & PR + push + rel            & 26             & 4                               \\
issue\_comment        & 2,650         & 1.2         & schedule only              & 25             & 3                               \\
workflow\_run         & 2,174         & 1.0         & PR only                    & 21             & 3                               \\
issues                & 1,743         & 0.8         & PR + push + sched + WD     & 18             & 2                               \\
workflow\_dispatch    & 1,654         & 0.8         & PR + push + WD             & 15             & 2                               \\
release               & 616           & 0.3         & dyn + PR + push + sched    & 15             & 2                               \\ 
\bottomrule
\end{tabular}
\footnotesize
\\sched = schedule; dyn = dynamic; rel = release; WD = workflow\_dispatch
\vspace{-5mm}
\end{table}

\noindent
\textbf{Development-Centric Trigger Events:} 
we observe that 
\texttt{push} (36.0\% of runs, primary in 332 repositories) and \texttt{pull\_request} (35.8\% of runs, primary in 250 repositories) account for 71.8\% of all runs. 
The most common repository strategy combines both triggers: 238 repositories (32\% of all repositories) use the \texttt{push} + \texttt{pull\_request} combination, enabling coverage of both direct commits and PRs (Table~\ref{tab:trigger-event-details}-b). 

\noindent
\textbf{Automation Focused Trigger Events:} 
we see that teams selectively adopt automation-focused trigger events. \texttt{schedule} triggers account for 18.6\% of runs and serve as the primary trigger in 108 repositories, similar to our observations in high-intensity utilizers like \texttt{aws-cloudformation/\allowbreak cfn-lint-visual-studio-code} with 878 scheduled runs (out of 1000) for automated schema updates. 
\texttt{dynamic} triggers (3.0\% of runs) represent automated services like Dependabot and security scans, serving as primary triggers in 22 repositories but appearing across more repositories as supplementary automation (Table~\ref{tab:sampling_classification}-b).

\begin{figure}[!b]
\centering
\vspace{-5mm}
\includegraphics[width=0.48\textwidth]{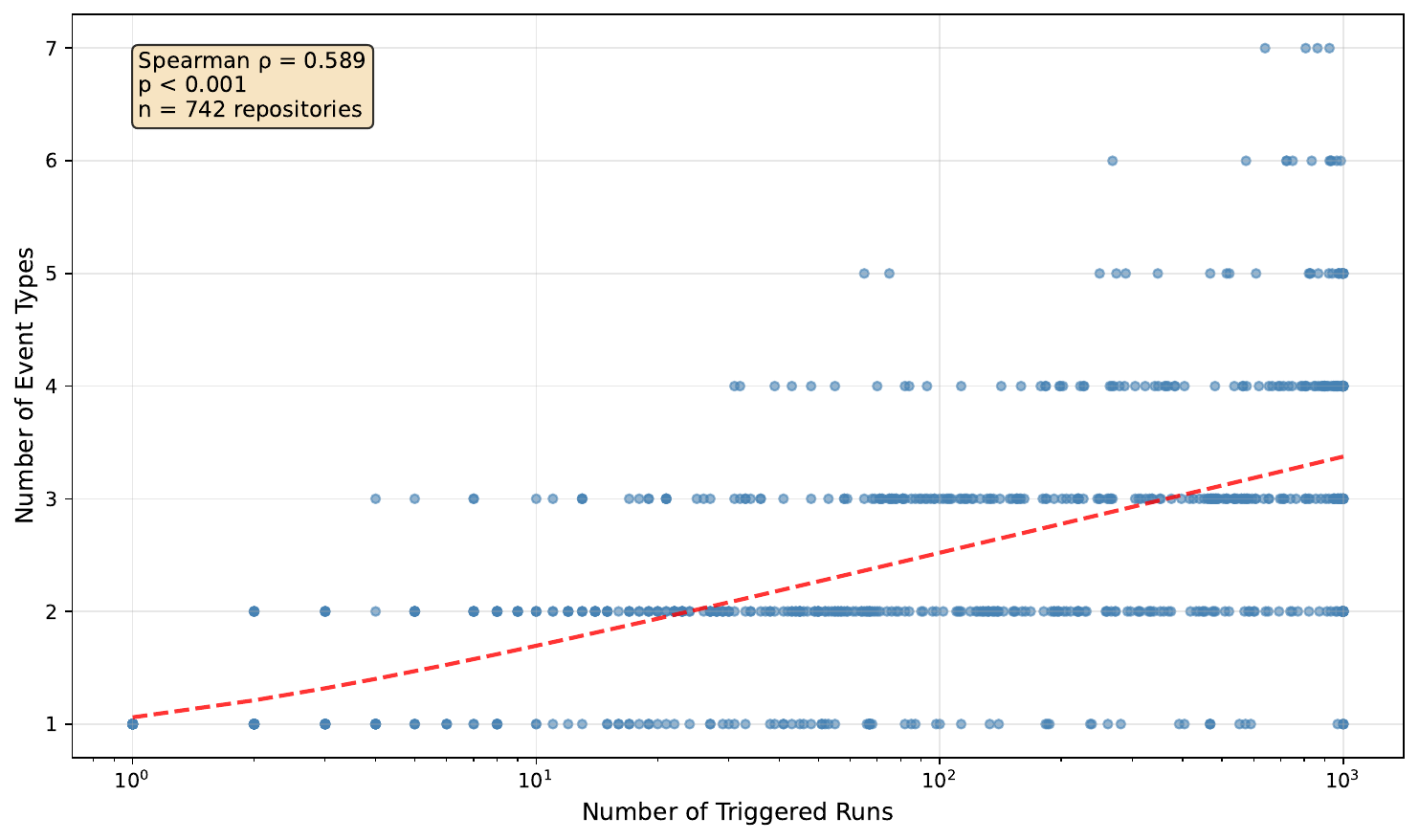}
\vspace{-2mm}
\caption{Trigger event diversity increasing as the \# of runs increase.}
\vspace{-1mm}
\label{fig:trigger-event-diversity}
\end{figure}

\noindent
\textbf{Repository Trigger Diversity Patterns:} 
repositories demonstrate variety in trigger event usage, with an average of 2.5 events used per repository. 
We also observe that an increased event diversity in a repository correlates with 
having more run records (Figure~\ref{fig:trigger-event-diversity}).
However, trigger concentration reveals strategic focus: 284 repositories (38.3\%) concentrate more than 80\% of their runs in a single trigger event. 
At the other extreme, repositories like \texttt{BeyondDimension/SteamTools} utilize more trigger event diversity (791 \texttt{push} + 98 \texttt{pull\_request} + 69 \texttt{create} + 42 \texttt{delete}). 

The trigger pattern analysis reinforces our intensity findings: higher-usage repositories ($\geq$100 runs) employ diverse trigger combinations, averaging 3.04 unique events, while low-usage repositories (<100 runs) average 1.90 trigger event types. 
Due to the non-linear relationship visible in Figure~\ref{fig:trigger-event-diversity}, we used Spearman's rank correlation~\cite{cohen2013applied}, which reveals a strong positive correlation between trigger diversity and run count ($\rho = 0.589$, $p < 0.001$). This pattern is consistent: 33.7\% of low-usage repositories rely on single triggers compared to 6.3\% of higher-usage repositories.

\colorlet{shadecolor}{gray!20}
\begin{shaded}
\textbf{RQ1 Summary:} GitHub Actions utilization follows distinct patterns: \textbf{High-usage repositories} 
have low failure rates (2.7--20.5\%) with more diverse trigger combinations, while \textbf{low-usage repositories} 
exhibit extreme variability (0--86.7\% failure rates) with predominantly single-trigger strategies (33.7\% use one trigger event). 
\textbf{Development-centric events triggers} dominate: \texttt{push} (36.0\%) and \texttt{pull\_request} (35.8\%) account for 71.8\% of runs. The negative correlation between run count and failure rates combined with strong correlation between trigger diversity and usage intensity indicates that repositories with sustained workflow usage develop more stable (less failure rate) and comprehensive workflow adoption (diverse trigger events).
\end{shaded}

\begin{figure*}[htbp]
\centering
\includegraphics[width=1\textwidth]{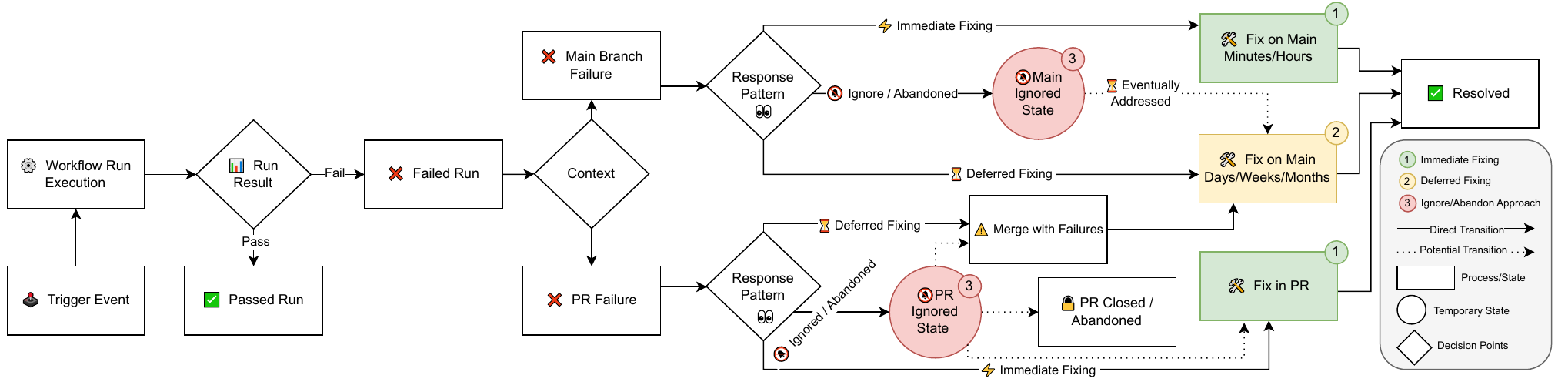}
\vspace{-4.5mm}
\caption{Developer Response Patterns to Workflow Failures in PR and Main Branch Contexts}
\label{fig:failure-reaction-model}
\vspace{-3.5mm}
\end{figure*}

\section{\textbf{RQ2: How do developers react to GitHub Actions workflow failures?}}\label{section:RQ2}

After understanding the broader workflow utilization of our large sample of 765 repositories, we investigate in more depth how developers react to workflow failures.
As this focuses on understanding the context and unique situations of the project, we opt for a qualitative manual analysis.

Our analysis of workflow failure responses across our 21 sampled repositories (Section \ref{section:methodology}) reveals distinct patterns in how development teams handle GitHub Actions failures. The patterns are not mutually exclusive;  projects can exhibit multiple response behaviors. 
Four repositories in our sampled dataset experienced no failures, so no reactions were observed for them.

We distinguished workflow failures (1) in pull requests (analyzed through PR comments and commits) and (2) on the main branch (tracked through subsequent pushes and timing), as these contexts present different response opportunities and constraints. Figure~\ref{fig:failure-reaction-model} illustrates  how our observed failure reaction patterns occur.
We now discuss each pattern. 

\subsection{\textbf{Fix-Oriented Response Patterns}}
\noindent
\subsubsection{\textbf{Immediate Fixing (13 repositories)}} 
When developers address workflow failures directly within pull requests before merging, or by pushing commits directly to the main branch \emph{immediately} within minutes or hours after (less than 24 hours).

\noindent
\textbf{Examples: }
\repo{webextension-toolbox/webextension-toolbox} shows collaborative fixing: when a contributor's PR fails lint checks, the maintainer (reviewer of the PR) fixes it in the same PR (\#866) or guides the contributor to fix it before merging (PR \#890). In \repo{trusttoken/contracts-pre22} the contributor makes multiple commits to resolve failed run issues before merging (PR \#1264).
In \repo{data-driven-forms/react-forms} the contributor gets feedback from the PR reviewer and keeps pushing fix changes until all check runs pass before a merge (PR \#1492). 
Developers also show immediate fixing behavior when directly pushing to the main branch of a repository. For example, in \repo{ai/size-limit} when there is a commit with failed runs on the main branch, we see cases of immediate fixing in the subsequent commit(s) happening \textit{shortly} after, e.g., commit \href{https://github.com/ai/size-limit/commit/058b1f132c4e51272e94e9d3650e480b3e0d6851}{058b1f1} workflow runs pass 3 minutes after a commit with a failed run. 
In \repo{BeyondDimension/SteamTools} we observe instances of fixing on the same day (commit \href{https://github.com/BeyondDimension/SteamTools/commit/078d4623f9f063c7ffbe163681bf9837cffc63b3}{078d462}, and commit \href{https://github.com/BeyondDimension/SteamTools/commit/36339154dc6e5137217aa1a923a1240fe8c001f6}{3633915}). 

\repo{iouAkira/someDockerfile} shows the same behavior but with a different goal: the user keeps pushing changes until the run for Docker image builds passes, effectively treating GitHub Actions as a deployment service to push images to the Docker Hub. \repo{RailsEventStore/rails\_event\_store} demonstrates an explicit culture of keeping all runs passing where failed commits are followed by immediate fixing commits 
(we did not observe a deferred fix approach for this repository).
\repo{aws-cloudformation/cfn-lint-visual-studio-code} shows a culture of ensuring workflows run successfully.
The few run failures that we observed
were fixed during the related PR.

\noindent
\textbf{Summary:} immediate fixing, the most prevalent pattern observed in 76\% of repositories with run failures, represents a proactive approach where teams resolve  workflow failures within minutes to hours. This pattern manifests both in the pull-based development model~\cite{gousios2014pullbased,gousiosICSE2015,veenMSR2025} and in projects that directly commit to the main branch 
(Figure~\ref{fig:failure-reaction-model} -- \circled{1}).

\subsubsection{\textbf{Deferred Fixing (6 repositories)}} Occurs when teams accept temporary failures in their main branch and address them in subsequent commits. We distinguish between immediate and deferred fixes based on timing. If the failure occurs on the main branch and the  commit that resolves the workflow failure happens days ($\geq$ 24 hours) later, rather than within minutes or hours, we consider it a deferred fix.

\noindent
\textbf{Examples:} in \repo{libcpr/cpr} (PR \#1170) a maintainer states \textit{``The CI failures are expected and I will fix them soon on master,''} showing a deferred fixing approach to address failures post-merge rather than blocking development velocity (later fixed in PR \#1173). \repo{komamitsu/fluency} exhibits deferred fixing where a Dependabot version bump caused CI failures 
(commit \href{https://github.com/komamitsu/fluency/commit/c870a74c4eeaab5247ae1280efdc98e01b72aa85}{c870a74}), but the maintainer addressed the issue almost 2 months later through a separate PR (\#915).

In \repo{BeyondDimension/SteamTools} we observed multiple deferred fixing examples: PR \#3469 merged with failed runs and was fixed 2 days later (commit \href{https://github.com/BeyondDimension/SteamTools/commit/7b9ba5e8a0108a77ad955a3fcffc3050517bb6f7}{7b9ba5e}), and a prolonged CI failure period of two months (between commits \href{https://github.com/BeyondDimension/SteamTools/commit/150f93e7b7cb9d0e236055a1639843b65e0a8796}{150f93e} and \href{https://github.com/BeyondDimension/SteamTools/commit/c087a02303d8debc94a7f55c6b0fd48991013986}{c087a02}) 
caused by an UI framework upgrade,
 was eventually resolved through workflow configuration updates. 

The deferred fixing approach spans varying timeframes, from days to months, reflecting different tolerances for temporary failures. \repo{yandex-cloud/serverless-plugin} exhibits extended delays with NodeJS version-related failures taking almost a month after the first failure to be addressed through dependency upgrades, showing tolerance for dependency-related issues. \repo{libcpr/cpr} demonstrates a deferred fixing with a 4-month gap between failed and successful workflow runs on master (between commit \href{https://github.com/libcpr/cpr/commit/372b54c5d91d82b41b9e93ac2ec13bd9d2c3d0a0 }{372b54c} and commit \href{https://github.com/libcpr/cpr/commit/a9466f70156ae159711aa9bae868a986504b2506}{a9466f7}), indicating acceptance of non-critical deployment failures until the next release cycle.

\noindent
\textbf{Summary:} the deferred fixing approach reflects tolerance for temporary failures, with resolution timeframes spanning days to months. 
Teams employing this pattern appear to distinguish critical path failures requiring immediate attention and issues that can be addressed during planned maintenance.
This prioritizes development velocity over passing workflow runs, accepting temporary failed runs in exchange for unblocked progress (Figure~\ref{fig:failure-reaction-model} -- \circled{2}). 

\subsubsection{\textbf{Ignore/Abandon Approach (4 repositories)}} Describes workflow failures that receive no addressing action,
representing a temporary state where teams either deliberately accept certain failures, face resource constraints preventing  attention, or ultimately remove (or disable) failed workflows.
In PRs, failures are ignored if they receive no reaction comments or fixing commits, or if the PR is closed/abandoned without resolution. On the main branch, failures remain ignored if they persist through subsequent commits without addressing action (Figure~\ref{fig:failure-reaction-model} -- \circled{3}). Our study window is a snapshot: ignored failures may later transition to the deferred fixing pattern. 

\noindent
\textbf{Examples:} several repositories demonstrate explicit decisions to ignore certain workflow failures with clear justifications. 
\repo{boutproject/BOUT-dev} demonstrates platform-specific ignoring where the reviewer explicitly states \textit{``LGTM, fedora failure is unrelated''} when merging PRs with Fedora platform CI failures (\#3029), and also states \textit{``I am happy to remove the coverage run. I have not looked at the results in years...''} showing ignoring test coverage failures (\#3067).
This indicates an approach of dismissing environment-specific 
issues deemed non-critical. Similarly, in \repo{data-driven-forms/react-forms} PR \#1449 was merged while CodeCov coverage check failed, indicating tolerance for specific quality assurance metric failures. Another example is \repo{ngageoint/geopackage-js}, where sometimes the last commit remains failing for a while (e.g., commit \href{https://github.com/ngageoint/geopackage-js/commit/3981bf07a68bb3df893df58231f4c0fcfa9acd65}{3981bf0}), potentially representing a conscious trade-off between CI/CD hygiene and development velocity.

Complete abandonment of workflow failures occurs when teams determine to remove or disable the workflow. \repo{BeyondDimension/SteamTools} provides a clear example
with marking \texttt{dotnet.yml} and \texttt{xamarin.yml} as \emph{disabled} after persistent failures.
The repository also shows workflows abandonment: the \emph{publish} workflow never ran successfully after commit \href{https://github.com/BeyondDimension/SteamTools/commit/b40dac00b971732b2dcd7de8ef1b47d7f847fbe2}{b40dac0}, with the team stopping attempts to fix (till the time of our study). 
\repo{trusttoken/contracts-pre22} demonstrates migration from GHA to CircleCI (abandoning GHA) and then again abandoning CircleCI by removing the workflow; also disabling their  Coveralls test workflow.
\repo{iouAkira/someDockerfile} has 13 different workflow configurations but only 2 with run records, suggesting that most workflows were effectively disabled after initial setup. 

\noindent
\textbf{Summary:} the ignore/abandon approach reflects three distinct scenarios: deliberate acceptance of failed runs, resource constraints that prevent immediate attention to failed runs, or removal of workflows causing failures. Teams may ultimately resolve ignored failures through three pathways: transitioning to immediate or deferred fixing, continuing to leave them unaddressed, or removing failing workflows entirely. The temporary nature of this classification highlights that ignored failures represent a point-in-time observation rather than a permanent team strategy, with resolution approaches potentially evolving as project priorities or resources change.

\colorlet{shadecolor}{gray!20}
\begin{shaded}
\textbf{RQ2 Summary:} our analysis of 21 repositories reveals three distinct response patterns: immediate fixing (13 of repositories) where teams resolve failures within minutes to hours, reflecting a commitment to passing continuous integration; deferred fixing (6 repositories) where teams strategically tolerate temporary failures, distinguishing between critical issues and those addressable in the future; and ignore/abandon (4 repositories) where failures receive no immediate action due to deliberate acceptance, potential resource constraints, or eventual workflow removal. These non-mutually exclusive patterns demonstrate that teams make nuanced prioritization decisions potentially based on failure context, project constraints, and strategic considerations.
\end{shaded}

\section{\textbf{RQ3: What relationships exist between project characteristics and GitHub Actions utilization patterns? }}
\label{section:RQ3}

Understanding how project characteristics influence GitHub Actions utilization is fundamental for developing context-aware CI/CD research and practice~\cite{khatamiSCAM2024}. 
In this section, we combine the insights from our quantitative analysis (Section~\ref{section:RQ1}) and in-depth qualitative analysis (Section~\ref{section:RQ2}).
We examine relationships between project characteristics (popularity, development style, team size, workflow evolution, and activity level) relate to utilization patterns (RQ1) and failure reactions (RQ2).
Given our limited sample size from our qualitative analysis, we deliberately emphasize these as hypotheses that represent a research agenda for future quantitative validation in large-scale mining studies.

\begin{table}[t]
\centering
\caption{Workflow utilization metrics (from RQ1).}
\vspace{-2mm}
\scriptsize
\begin{tabular}{llcccc}
\toprule
\textbf{Characteristic} & \textbf{Category} & \textbf{N} & \textbf{Runs} & \textbf{Fail \%} & \textbf{Div.} \\
\midrule
\textbf{Activity}   & High        & 7  & 620.9 & 28.2 & 3.1 \\
                    & Medium      & 7  & 258.0 & 7.8  & 2.4 \\
                    & Low         & 7  & 143.0 & 23.2 & 2.6 \\
\midrule
\textbf{Workflow Ev.}   & High        & 5  & 827.0 & 30.3 & 3.6 \\
                    & Moderate    & 5  & 459.0 & 22.8 & 3.4 \\
                    & Minimal     & 7  & 81.3  & 19.5 & 2.0 \\
                    & No          & 4  & 38.5  & 3.2  & 2.0 \\
\midrule
\textbf{Team Size}  & Solo        & 3  & 339.0 & 29.8 & 2.3 \\
                    & 1 maint.    & 10 & 282.6 & 23.9 & 2.8 \\
                    & Multi maint.& 8  & 413.8 & 10.8 & 2.8 \\
\midrule
\textbf{Dev. Style} & Direct push & 2  & 8.5   & 43.3 & 1.5 \\
                    & Mixed       & 11 & 343.9 & 22.7 & 2.7 \\
                    & Pull-based  & 8  & 419.1 & 9.7  & 3.0 \\
                    
\midrule           
\textbf{Popularity} & High & 10 & 455.5 & 20.7 & 2.8 \\
                    & Low  & 11 & 236.2 & 18.9 & 2.6 \\
\bottomrule
\end{tabular}
\label{tab:relations-with-workflow-utilization}
\footnotesize
\\Runs = average workflow runs; Fail \% = average failure rate; Div. = average trigger event diversity
\vspace{-5mm}
\end{table}

\subsection{Project Characteristics and Workflow Utilization Patterns}

Table~\ref{tab:relations-with-workflow-utilization} presents workflow utilization metrics grouped by project characteristics. 
These observed patterns inform our hypotheses (H1-H3) for future validation.

\noindent
\textbf{Activity Level:} high-activity projects show the most intensive workflow usage with 620.9 average runs and the highest event diversity (3.1), but also experience higher failure rates (28.2\%). Medium-activity projects achieve the lowest failure rates (7.8\%) despite moderate usage (258 runs). Low-activity projects show limited workflow adoption (143 runs) with surprisingly higher failure rates (23.2\%), potentially due to experimental or one-off task usage (e.g. \repo{iouAkira/someDockerfile}).

\noindent
\textbf{Workflow Evolution:} projects with high workflow evolution demonstrate the most intensive usage patterns (827 runs) and highest event diversity (3.6), but also highest failure rates (30.3\%). Projects with no workflow evolution show minimal usage (38.5 runs) with the lowest failure rates (3.2\%). 
Moderate evolution balances usage (459 runs) with failure rates (22.8\%) closer to the overall average (19.7\%).

\noindent
\textbf{Team Size:} multi-maintainer projects have the highest run counts (413.8), yet lowest failure rates (10.8\%). Solo projects demonstrate moderate usage (339 runs) but high failure rates (29.8\%), while single-maintainer projects show the lowest usage (282.6 runs) with intermediate failure rates (23.9\%). Event diversity is consistent across team size (2.3-2.8), suggesting trigger complexity is independent of team structure.

\noindent
\textbf{Development Style:} pull-based development has the highest workflow usage (419.1 runs) and event diversity (3) with the lowest failure rates (9.7\%). Direct push approaches show minimal workflow adoption (8.5 runs) with high failure rates (43.3\%). Mixed approaches balance moderate usage (343.9 runs) with moderate failure rates (22.7\%).

\noindent
\textbf{Popularity:} high-popularity projects show nearly double the workflow usage of low-popularity projects (455.5 vs. 236.2 runs) with comparable failure rates (20.7\% vs. 18.9\%), suggesting that project popularity may drive workflow adoption intensity. Event diversity shows minimal variation (2.8 vs. 2.6), indicating that trigger patterns are consistent regardless of  popularity.

\begin{shaded}
\textbf{Hypothesis H1:} 
pull-based development is linked to lower workflow failure rates compared to direct push approaches.

\smallskip
\textbf{Hypothesis H2:} projects with higher workflow evolution intensity experience higher failure rates, suggesting that frequent configuration changes are linked to increased workflow failures.


\smallskip
\textbf{Hypothesis H3:} team size is negatively linked to workflow failure rates, with projects having multiple maintainers showing lower failure rates than solo projects, and single-maintainer projects falling between these extremes.

\end{shaded}

\begin{table}[t]
\centering
\caption{Workflow failure reaction patterns (from RQ2)}
\scriptsize
\vspace{-2mm}
\begin{tabular}{lcccc}
\toprule
\textbf{Characteristic} & \textbf{Total} & \textbf{Immediate} & \textbf{Deferred} & \textbf{Ignore} \\
\textbf{Categories} & \textbf{} & \textbf{Fixing (\%)} & \textbf{Fixing (\%)} & \textbf{/Abandon (\%)} \\
\midrule
\multicolumn{5}{l}{\textbf{Activity Level}} \\
High   & 7 & 5 (71.4) & 4 (57.1) & 4 (57.1) \\
Medium & 7 & 4 (57.1) & 2 (28.6) & 0 (0.0) \\
Low    & 7 & 4 (57.1) & 0 (0.0)  & 0 (0.0) \\
\midrule
\multicolumn{5}{l}{\textbf{Workflow Evolution}} \\
High     & 5 & 3 (60.0) & 4 (80.0) & 4 (80.0) \\
Moderate & 5 & 4 (80.0) & 0 (0.0)  & 0 (0.0) \\
Minimal  & 7 & 4 (57.1) & 1 (14.3) & 0 (0.0) \\
No       & 4 & 2 (50.0) & 1 (25.0) & 0 (0.0) \\
\midrule
\multicolumn{5}{l}{\textbf{Team Size}} \\
Solo project         & 3  & 2 (66.7) & 0 (0.0)  & 0 (0.0) \\
1 maintainer         & 10 & 5 (50.0) & 3 (30.0) & 2 (20.0) \\
Multiple maintainers & 8  & 6 (75.0) & 3 (37.5) & 2 (25.0) \\
\midrule
\multicolumn{5}{l}{\textbf{Development Style}} \\
Direct push   & 2  & 1 (50.0) & 0 (0.0)  & 0 (0.0) \\
Mixed         & 11 & 8 (72.7) & 2 (18.2) & 2 (18.2) \\
Pull-based    & 8  & 4 (50.0) & 4 (50.0) & 2 (25.0) \\
\midrule
\multicolumn{5}{l}{\textbf{Popularity}} \\
High & 10 & 6 (60.0) & 3 (30.0) & 3 (30.0) \\
Low  & 11 & 7 (63.6) & 3 (27.3) & 1 (9.1) \\
\bottomrule
\end{tabular}
\\Four repositories had no failures. Workflow failure reaction patterns are not mutually exclusive, so percentages do not add up to 100\%.
\vspace{-9mm}
\label{tab:relations-with-fixing-approaches}
\end{table}

\subsection{Project Characteristics and Workflow Failure Reaction Patterns}

Table~\ref{tab:relations-with-fixing-approaches} summarizes how projects with different characteristics react to workflow failures.
These observed patterns inform our hypotheses (H4 and H5) for future validation.

\noindent
\textbf{Activity Level:} projects with high activity levels demonstrate the most diverse failure reaction strategies, with 71.4\% employing immediate fixes, and 57.1\% using both deferred fixes and ignore approaches. This suggests that actively developed projects have more flexibility in their failure response strategies. In contrast, projects with low activity exclusively use immediate fixes when they address failures, never showing deferred or ignore fixing among our studied projects, potentially reflecting their more focused scope and simpler workflows. 

\noindent
\textbf{Workflow Evolution:} a pattern emerges where projects with frequent workflow configuration updates show the highest rates of both deferred fixing (80\%) and ignoring failures (80\%), while projects with moderate evolution demonstrate the opposite behavior: 80\% immediate fixing with no deferred or ignore strategies. This counterintuitive finding suggests that frequent workflow changes may create technical debt that teams manage through selective attention to failures.

\noindent
\textbf{Team Size:} projects with multiple maintainers show a higher rate of immediate fixes (75\%), compared to single-maintainer projects (50\%). Solo projects never employ deferred or ignore strategies.
This suggests that projects with multiple maintainers are more likely to have diverse failure response strategies, potentially reflecting prioritization,  distributed responsibilities, and project complexity.

\noindent
\textbf{Development Style:} development approaches combining PRs with direct pushes (mixed) are more likely to do immediate fixes 
(72.7\%), while pull-based development shows equal distribution between immediate (50.0\%) and deferred fixes (50.0\%). 

\noindent
\textbf{Popularity:} repository popularity shows little to no relation with failure reaction patterns, with both high and low popularity projects demonstrating similar distributions across approaches. This indicates that reactions to workflow failures are independent of repository popularity.


\colorlet{shadecolor}{gray!20}
\begin{shaded}

\textbf{Hypothesis H4:} projects with high workflow evolution rates are more likely to employ selective failure fixing strategies (deferred and ignore approaches), potentially indicating that rapid configuration changes create technical debt that teams should  prioritize selectively.

\smallskip
\textbf{Hypothesis H5:} projects with multiple maintainers are more likely to employ immediate fixing strategies than single-maintainer projects.
\end{shaded}

\section{Discussion}\label{section:discussion}

\subsection{Collaborative vs. Direct Fixing in Workflow Failures}
Our observations reveal that while contributors may resolve workflow failures independently, maintainer involvement creates a duality in fixing approaches: maintainers either (1) guide external contributors to fix issues themselves, or (2) directly resolve failures within the PR. This pattern can reflect decisions about knowledge transfer when maintainers choose to intervene in case of workflow run failure in a pull request. When maintainers provide guidance, failures become learning opportunities for contributors to understand project context~\cite{liHMI2022}. When maintainers fix directly, development velocity is preserved but the knowledge sharing value is lost, potentially affecting contributor retention. 

\subsection{Local Optimization vs. 
Collective Costs 
}\label{dis:collective-costs}

The deferred fixing approach that we observe  in nearly 1/3 of our sampled repositories, reveals an asymmetry: while teams may make locally rational decisions to defer fixes, this can impose hidden costs on the 
development community.

We saw teams may make calculated decisions about failure priority, treating certain workflow issues as manageable technical debt, e.g., deferred fixing. Maintainers may reasonably defer fixes for infrastructure issues or dependency problems that do not block core development workflows. However, this local optimization can create problematic downstream effects. When subsequent contributors encounter failing builds, they may assume their changes caused the failure, leading to wasted debugging effort. 
This \emph{failure contamination} represents an asymmetric cost distribution where one team's deferred fix may generate time waste elsewhere. 

Additionally, normalization of failing workflows, may diminish contributor confidence in GHA workflows and reduce sensitivity to genuine new issues.

\subsection{The Configuration-Usage Gap}
Our approach of examining workflow run data rather than static configuration files, reveals methodological considerations for GHA workflows research. While most  previous studies analyzing GHA adoption focused on workflow configurations (Section~\ref{section:background}), our study uncovered cases where configurations did not reflect actual adoption: disabled workflows (\texttt{nextcloud/ocsms}, with CodeQL scans disabled and no runs in years), unused configurations (\texttt{iouAkira/someDockerfile} with 13 configurations but only 2 active), and abandoned workflows. This suggests that configuration-based analyses may overestimate actual workflow adoption and may miss the nuanced reality. 

\subsection{Ideas for Enhancing GitHub Actions}
Our findings illuminate areas where the current GHA platform can better support distributed development coordination:

\noindent
\textbf{Intelligent Failure Context}: providing automated failure attribution and historical context to help contributors distinguish between inherited issues and newly-introduced problems. 

\noindent
\textbf{Adaptive Mentoring Support}: enhanced tooling could automatically generate project-specific guidance for common workflow failures, reducing maintainer mentoring burden while preserving contributor learning opportunities~\cite{bacchelli2013expectations}.

\noindent
\textbf{Usage-Aware Analytics}: moving beyond configuration to usage-based insights would help teams understand automation adoption patterns and identify underutilized workflows.

\subsection{Threats to Validity}
\textbf{Internal validity:} Subjective interpretation of developer behavior was mitigated through dual coding and negotiated agreement between two authors.
\noindent
\textbf{External validity:} Our 21 repositories, while diverse across key dimensions, may not represent all GHA usage patterns. We frame RQ3 findings as hypotheses requiring future large-scale validation. 
\noindent
\textbf{Construct validity:} Our observation of reactions may not fully reflect underlying team intentions. The 1000-run collection limit may shorten data for highly active projects. The thresholds used in RQ1 are subjective and for interpretation only.

\section{Conclusion and Future Work}\label{section:conclusion}
We examined GitHub Actions adoption through workflow execution data and developer behavior, analyzing 21 repositories to understand how teams utilize automation in practice.
Analyzing 765 repositories (RQ1) revealed that higher usage intensity correlates with lower failure rates: repositories with more runs have lower failure rates and use diverse triggers, while repositories with fewer runs show high variability with predominantly single-trigger strategies.
Building on these findings, we identified three distinct failure response patterns (RQ2): \emph{immediate fixing}, resolving failures within hours, \emph{deferred fixing}, tolerating temporary failures for days to months, and \emph{ignore/abandon} approaches, where failures receive no immediate action or lead to workflow removal. These patterns indicate teams make strategic prioritization decisions based on context.
Finally, our mixed-methods analysis, combining quantitative metrics from 21 repositories with qualitative examination of their contexts, yielded five hypotheses for future validation (RQ3): workflows with more changes show more selective fixing strategies; multiple maintainers enable higher immediate fixing rates; pull-based development shows lower workflow failure rates; workflows with more changes have higher failure rates; and larger teams experience fewer workflow failures.

In terms of future work, we propose to (1) perform a \emph{large-scale validation} across diverse repositories. 
Statistical analysis should establish the generalizability of relationships between project characteristics and workflow utilization patterns identified in our study. 
Additionally, (2) we need to better understand the cost-benefit relationship of different workflow strategies, e.g., the costs of a deferred fixing pattern.
Finally, (3) we intend to investigate AI-assisted workflow guidance systems, and intelligent failure attribution to reduce the cost of fixing workflow failures.

\section*{Data Availability}
All the data, data collection pipelines, and analyses are available in our replication package~\cite{khatamiSANERReplicationPackage}.

\begin{acks}
This research was partially funded by the Dutch science foundation NWO
through the Vici ``TestShift'' grant (No. VI.C.182.032).
\end{acks}

\balance
\bibliography{bibliography}

@inproceedings{vasilescuESECFSE2015, 
    author = {Vasilescu, Bogdan and Yu, Yue and Wang, Huaimin and Devanbu, Premkumar and Filkov, Vladimir},
    title = {Quality and productivity outcomes relating to continuous integration in {GitHub}}, 
    year = {2015}, 
    isbn = {9781450336758}, 
    publisher = {ACM}, 
    -address = {New York, NY, USA}, 
    -url = {https://doi.org/10.1145/2786805.2786850}, 
    -doi = {10.1145/2786805.2786850}, 
    booktitle = {Proceedings of the 2015 10th Joint Meeting on Foundations of Software Engineering (ESEC/FSE)}, 
    pages = {805--816}, 
    numpages = {12}, keywords = {Continuous integration, GitHub, pull requests}, 
    -location = {Bergamo, Italy}, 
    -series = {ESEC/FSE 2015} }

@article{liHMI2022,
author = {Li, Zhixing and Yu, Yue and Wang, Tao and Li, Shanshan and Wang, Huaimin},
title = {Opportunities and Challenges in Repeated Revisions to Pull-Requests: An Empirical Study},
year = {2022},
issue_date = {November 2022},
publisher = {ACM},
-address = {New York, NY, USA},
volume = {6},
number = {CSCW2},
-doi = {10.1145/3555208},
journal = {Proc. ACM Hum.-Comput. Interact.},
month = nov,
articleno = {317},
numpages = {35},
keywords = {code review, open source software, pull-request, repeated revision}
}

@misc{khatamiSANERReplicationPackage,
  author       = {Ali Khatami and Carolin Brandt and Andy Zaidman},
  title        = {Replication Package for ``{Beyond} the {YAML} {F}ile:
                   {U}nderstanding {R}eal-{W}orld {G}itHub {A}ctions {W}orkflow
                   {A}doption
                  },
  year         = 2026,
  publisher    = {Zenodo},
  doi          = {10.5281/zenodo.18258226},
  url          = {https://doi.org/10.5281/zenodo.18258226},
}

@MISC{Fowler2006,
  author = {M. Fowler and M. Foemmel},
  title = {Continuous integration},
  url = {https://tinyurl.com/ycbl2uhj}, 
  note = {[Online; accessed 29-May-2025]}}

@ARTICLE{elazharyTSE2022,
  author={Elazhary, Omar and Werner, Colin and Li, Ze Shi and Lowlind, Derek and Ernst, Neil A. and Storey, Margaret-Anne},
  journal={IEEE Transactions on Software Engineering}, 
  title={Uncovering the Benefits and Challenges of Continuous Integration Practices}, 
  year={2022},
  volume={48},
  number={7},
  pages={2570-2583},
  doi={10.1109/TSE.2021.3064953}}

@INPROCEEDINGS{johnsonICSE2013,
  author={Johnson, Brittany and Song, Yoonki and Murphy-Hill, Emerson and Bowdidge, Robert},
  booktitle={International Conference on Software Engineering (ICSE)}, 
  title={Why don't software developers use static analysis tools to find bugs?}, 
  year={2013},
  volume={},
  number={},
  pages={672--681},
  publisher={IEEE},
  keywords={Interviews;Computer bugs;Encoding;Software;Teamwork;Companies;Standards},
  -doi={10.1109/ICSE.2013.6606613}}

@inproceedings{bellerMSR2017a,
  author       = {Moritz Beller and
                  Georgios Gousios and
                  Andy Zaidman},
  -editor       = {Jes{\'{u}}s M. Gonz{\'{a}}lez{-}Barahona and
                  Abram Hindle and
                  Lin Tan},
  title        = {TravisTorrent: synthesizing Travis {CI} and GitHub for full-stack
                  research on continuous integration},
  booktitle    = {Proceedings of the 14th International Conference on Mining Software
                  Repositories ({MSR})}, 
  pages        = {447--450},
  publisher    = {{IEEE}},
  year         = {2017},
  -url          = {https://doi.org/10.1109/MSR.2017.24},
  -doi          = {10.1109/MSR.2017.24},
  timestamp    = {Thu, 23 Mar 2023 23:57:40 +0100},
  biburl       = {https://dblp.org/rec/conf/msr/BellerGZ17a.bib},
  bibsource    = {dblp computer science bibliography, https://dblp.org}
}

@inproceedings{bellerMSR2017,
  author       = {Moritz Beller and
                  Georgios Gousios and
                  Andy Zaidman},
  -editor       = {Jes{\'{u}}s M. Gonz{\'{a}}lez{-}Barahona and
                  Abram Hindle and
                  Lin Tan},
  title        = {Oops, my tests broke the build: an explorative analysis of Travis
                  {CI} with GitHub},
  booktitle    = {Proceedings of the 14th International Conference on Mining Software
                  Repositories ({MSR})},
  pages        = {356--367},
  publisher    = {{IEEE}},
  year         = {2017},
  -url          = {https://doi.org/10.1109/MSR.2017.62},
  -doi          = {10.1109/MSR.2017.62},
  timestamp    = {Thu, 23 Mar 2023 23:57:40 +0100},
  biburl       = {https://dblp.org/rec/conf/msr/BellerGZ17.bib},
  bibsource    = {dblp computer science bibliography, https://dblp.org}
}

@inproceedings{bholanathSANER2016,
  author       = {Moritz Beller and
                  Radjino Bholanath and
                  Shane McIntosh and
                  Andy Zaidman},
  title        = {Analyzing the State of Static Analysis: {A} Large-Scale Evaluation
                  in Open Source Software},
  booktitle    = {{IEEE} 23rd International Conference on Software Analysis, Evolution,
                  and Reengineering ({SANER})}, 
  pages        = {470--481},
  publisher    = {{IEEE}},
  year         = {2016},
  -url          = {https://doi.org/10.1109/SANER.2016.105},
  -doi          = {10.1109/SANER.2016.105},
  timestamp    = {Fri, 24 Mar 2023 00:04:45 +0100},
  biburl       = {https://dblp.org/rec/conf/wcre/BellerBMZ16.bib},
  bibsource    = {dblp computer science bibliography, https://dblp.org}
}

@inproceedings{veenMSR2025,
  author       = {Erik van der Veen and
                  Georgios Gousios and
                  Andy Zaidman},
  -editor       = {Massimiliano Di Penta and
                  Martin Pinzger and
                  Romain Robbes},
  title        = {Automatically Prioritizing Pull Requests},
  booktitle    = {12th {IEEE/ACM} Working Conference on Mining Software Repositories (MSR)},
  pages        = {357--361},
  publisher    = {{IEEE}},
  year         = {2015},
  -url          = {https://doi.org/10.1109/MSR.2015.40},
  -doi          = {10.1109/MSR.2015.40},
  timestamp    = {Thu, 23 Mar 2023 23:57:40 +0100},
  biburl       = {https://dblp.org/rec/conf/msr/VeenGZ15.bib},
  bibsource    = {dblp computer science bibliography, https://dblp.org}
}

@inproceedings{gousiosICSE2015,
  author       = {Georgios Gousios and
                  Andy Zaidman and
                  Margaret{-}Anne D. Storey and
                  Arie van Deursen},
  -editor       = {Antonia Bertolino and
                  Gerardo Canfora and
                  Sebastian G. Elbaum},
  title        = {Work Practices and Challenges in Pull-Based Development: The Integrator's
                  Perspective},
  booktitle    = {37th {IEEE/ACM} International Conference on Software Engineering (ICSE)},
  pages        = {358--368},
  publisher    = {{IEEE}},
  year         = {2015},
  -url          = {https://doi.org/10.1109/ICSE.2015.55},
  -doi          = {10.1109/ICSE.2015.55},
  timestamp    = {Thu, 23 Mar 2023 23:58:06 +0100},
  biburl       = {https://dblp.org/rec/conf/icse/GousiosZSD15.bib},
  bibsource    = {dblp computer science bibliography, https://dblp.org}
}

@article{besseyCACM2010,
  author       = {Al Bessey and
                  Ken Block and
                  Benjamin Chelf and
                  Andy Chou and
                  Bryan Fulton and
                  Seth Hallem and
                  Charles{-}Henri Gros and
                  Asya Kamsky and
                  Scott McPeak and
                  Dawson R. Engler},
  title        = {A few billion lines of code later: using static analysis to find bugs
                  in the real world},
  journal      = {Commun. {ACM}},
  volume       = {53},
  number       = {2},
  pages        = {66--75},
  year         = {2010},
  -url          = {https://doi.org/10.1145/1646353.1646374},
  -doi          = {10.1145/1646353.1646374},
  timestamp    = {Sun, 02 Jun 2019 20:48:59 +0200},
  biburl       = {https://dblp.org/rec/journals/cacm/BesseyBCCFHHKME10.bib},
  bibsource    = {dblp computer science bibliography, https://dblp.org}
}

@inproceedings{HiltonASE2016,
  author       = {Michael Hilton and
                  Timothy Tunnell and
                  Kai Huang and
                  Darko Marinov and
                  Danny Dig},
  -editor       = {David Lo and
                  Sven Apel and
                  Sarfraz Khurshid},
  title        = {Usage, costs, and benefits of continuous integration in open-source
                  projects},
  booktitle    = {Proceedings of the 31st {IEEE/ACM} International Conference on Automated
                  Software Engineering ({ASE})},
  pages        = {426--437},
  publisher    = {{ACM}},
  year         = {2016},
  -url          = {https://doi.org/10.1145/2970276.2970358},
  -doi          = {10.1145/2970276.2970358},
  timestamp    = {Sat, 30 Sep 2023 09:51:40 +0200},
  biburl       = {https://dblp.org/rec/conf/kbse/HiltonTHMD16.bib},
  bibsource    = {dblp computer science bibliography, https://dblp.org}
}

@book{glaser2017discovery,
  title={Discovery of grounded theory: Strategies for qualitative research},
  author={Glaser, Barney and Strauss, Anselm},
  year={2017},
  publisher={Routledge}
}

@article{DBLP:journals/iahe/GarrisonCKK06,
  author       = {D. Randy Garrison and
                  Martha Cleveland{-}Innes and
                  Marguerite Koole and
                  James Kappelman},
  title        = {Revisiting methodological issues in transcript analysis: Negotiated
                  coding and reliability},
  journal      = {Internet High. Educ.},
  volume       = {9},
  number       = {1},
  pages        = {1--8},
  year         = {2006},
  -url          = {https://doi.org/10.1016/j.iheduc.2005.11.001},
  -doi          = {10.1016/J.IHEDUC.2005.11.001},
  timestamp    = {Sun, 06 Oct 2024 21:28:41 +0200},
  biburl       = {https://dblp.org/rec/journals/iahe/GarrisonCKK06.bib},
  bibsource    = {dblp computer science bibliography, https://dblp.org}
}

@inproceedings{DBLP:conf/icsm/DecanMMG22,
  author       = {Alexandre Decan and
                  Tom Mens and
                  Pooya Rostami Mazrae and
                  Mehdi Golzadeh},
  title        = {On the Use of {GitHub} {Actions} in Software Development Repositories},
  booktitle    = {{IEEE} International Conference on Software Maintenance and Evolution,
                  ({ICSME})},
  pages        = {235--245},
  publisher    = {{IEEE}},
  year         = {2022},
  -url          = {https://doi.org/10.1109/ICSME55016.2022.00029},
  -doi          = {10.1109/ICSME55016.2022.00029},
  timestamp    = {Wed, 11 Jan 2023 16:58:31 +0100},
  biburl       = {https://dblp.org/rec/conf/icsm/DecanMMG22.bib},
  bibsource    = {dblp computer science bibliography, https://dblp.org}
}

@article{DBLP:journals/ese/WesselVGT23,
  author       = {Mairieli Wessel and
                  Joseph Vargovich and
                  Marco Aur{\'{e}}lio Gerosa and
                  Christoph Treude},
  title        = {GitHub Actions: The Impact on the Pull Request Process},
  journal      = {Empir. Softw. Eng.},
  volume       = {28},
  number       = {6},
  pages        = {131},
  year         = {2023},
  -url          = {https://doi.org/10.1007/s10664-023-10369-w},
  -doi          = {10.1007/S10664-023-10369-W},
  timestamp    = {Wed, 01 Nov 2023 08:59:33 +0100},
  biburl       = {https://dblp.org/rec/journals/ese/WesselVGT23.bib},
  bibsource    = {dblp computer science bibliography, https://dblp.org}
}

@inproceedings{DBLP:conf/icse/BouzeniaP24,
  author       = {Islem Bouzenia and
                  Michael Pradel},
  title        = {Resource Usage and Optimization Opportunities in Workflows of {GitHub}
                  {Actions}},
  booktitle    = {Proceedings of the 46th {IEEE/ACM} International Conference on Software
                  Engineering ({ICSE})}, 
  pages        = {25:1--25:12},
  publisher    = {{ACM}},
  year         = {2024},
  -url          = {https://doi.org/10.1145/3597503.3623303},
  -doi          = {10.1145/3597503.3623303},
  timestamp    = {Mon, 24 Jun 2024 15:20:25 +0200},
  biburl       = {https://dblp.org/rec/conf/icse/BouzeniaP24.bib},
  bibsource    = {dblp computer science bibliography, https://dblp.org}
}

@INPROCEEDINGS{khatamiSCAM2024,
  author={Khatami, Ali and Brandt, Carolin and Zaidman, Andy},
  booktitle={2024 IEEE International Conference on Source Code Analysis and Manipulation (SCAM)}, 
  title={Software Quality Assurance Analytics: Enabling Software Engineers to Reflect on {QA} Practices}, 
  year={2024},
  volume={},
  number={},
  pages={189-200},
  keywords={Software testing;Quality assurance;Reviews;Source coding;Prototypes;Software quality;Static analysis;Software systems;Software;Reflection;Software Quality Assurance;Software Analytics;Empirical Software Engineering;Software Testing;Code Review;Automation Workflows},
  -doi={10.1109/SCAM63643.2024.00027}}

@inproceedings{delicheh2023preliminary,
  title={A Preliminary Study of {GitHub} {Actions} Dependencies},
  author={Delicheh, Hassan Onsori and Decan, Alexandre and Mens, Tom},
  booktitle={SATToSE},
  pages={66--77},
  year={2023}
}

@inproceedings{golzadeh2022rise,
  title={On the rise and fall of {CI} services in {GitHub}},
  author={Golzadeh, Mehdi and Decan, Alexandre and Mens, Tom},
  booktitle={2022 IEEE International Conference on Software Analysis, Evolution and Reengineering (SANER)},
  pages={662--672},
  year={2022},
  organization={IEEE}
}

@article{rostami2023usage,
  title={On the usage, co-usage and migration of {CI/CD} tools: A qualitative analysis},
  author={Rostami Mazrae, Pooya and Mens, Tom and Golzadeh, Mehdi and Decan, Alexandre},
  journal={Empirical Software Engineering},
  volume={28},
  number={2},
  pages={52},
  year={2023},
  publisher={Springer}
}

@inproceedings{chen2021let,
  title={Let's supercharge the workflows: An empirical study of {GitHub} {Actions}},
  author={Chen, Tingting and Zhang, Yang and Chen, Shu and Wang, Tao and Wu, Yiwen},
  booktitle={2021 IEEE 21st International Conference on Software Quality, Reliability and Security Companion (QRS-C)},
  pages={01--10},
  year={2021},
  organization={IEEE}
}

@inproceedings{ayala2023empirical,
  title={An empirical study on workflows and security policies in popular github repositories},
  author={Ayala, Jessy and Garcia, Joshua},
  booktitle={2023 IEEE/ACM 1st International Workshop on Software Vulnerability (SVM)},
  pages={6--9},
  year={2023},
  organization={IEEE}
}

@inproceedings{kinsman2021software,
  title={How do software developers use github actions to automate their workflows?},
  author={Kinsman, Timothy and Wessel, Mairieli and Gerosa, Marco A and Treude, Christoph},
  booktitle={2021 IEEE/ACM 18th International Conference on Mining Software Repositories (MSR)},
  pages={420--431},
  year={2021},
  organization={IEEE}
}

@article{decan2023outdatedness,
  title={On the outdatedness of workflows in the {GitHub} {Actions} ecosystem},
  author={Decan, Alexandre and Mens, Tom and Delicheh, Hassan Onsori},
  journal={Journal of Systems and Software},
  volume={206},
  pages={111827},
  year={2023},
  publisher={Elsevier}
}

@inproceedings{rostami2024gawd,
  title={gawd: A differencing tool for GitHub Actions workflows},
  author={Rostami Mazrae, Pooya and Decan, Alexandre and Mens, Tom},
  booktitle={Proceedings of the 21st International Conference on Mining Software Repositories},
  pages={682--686},
  year={2024}
}

@inproceedings{valenzuela2022evolution,
  title={Evolution of github action workflows},
  author={Valenzuela-Toledo, Pablo and Bergel, Alexandre},
  booktitle={2022 IEEE International Conference on Software Analysis, Evolution and Reengineering (SANER)},
  pages={123--127},
  year={2022},
  organization={IEEE}
}

@inproceedings{benedetti2022automatic,
  title={Automatic security assessment of {GitHub} {Actions} workflows},
  author={Benedetti, Giacomo and Verderame, Luca and Merlo, Alessio},
  booktitle={Proceedings of the 2022 ACM Workshop on Software Supply Chain Offensive Research and Ecosystem Defenses},
  pages={37--45},
  year={2022}
}

@article{klotinsEMSE2022,
  author = {Klotins, Eriks and Gorschek, Tony and  Sundelin, Katarina and Falk, Erik},
  year = {2022},
  title = {Towards cost-benefit evaluation for continuous software engineering activities},
  journal = {Empirical Software Engineering},
  volume = {157},
  issue = {6},
  -doi = {10.1007/s10664-022-10191-w}
}

@article{santosEMSE2025,
author = {Santos, Jadson and da Costa, Daniel Alencar and McIntosh, Shane and Kulesza, Uir\'{a}},
title = {On the need to monitor continuous integration practices},
year = {2025},
issue_date = {Aug 2025},
publisher = {Kluwer Academic Publishers},
-address = {USA},
volume = {30},
number = {5},
issn = {1382-3256},
-url = {https://doi.org/10.1007/s10664-025-10682-6},
-doi = {10.1007/s10664-025-10682-6},
abstract = {One of the crucial activities in software development is monitoring. It plays a vital role in verifying the proper implementation of processes, the identification of errors, and the discovery of opportunities for improvement. Continuous Integration (CI) encompasses a set of widely adopted practices that enhance software development. However, there are indications that developers may not adequately monitor CI practices. Hence, this paper explores developers’ perceptions regarding monitoring CI practices. To achieve this, we first defined metrics associated with CI practices. Then, we conducted a series of analyses to evaluate how these metrics are perceived in projects or by their developers. As a starting point, we perform a Document Analysis to assess developers’ expressed need for practice monitoring in pull request comments generated by developers during the development process. After that, we conduct a survey of 28 developers from 121 open-source projects to understand the perception of the significance of monitoring seven CI practices in their projects. Finally, we triangulate the emergent themes from our survey by performing a second Document Analysis to understand the extent of monitoring features supported by existing CI services. Our key findings indicate that: 1) the most frequently mentioned CI practice during the development process is “Write automated developer tests”, which is associated with the metric “Coverage” (> 80\%), while “Don’t commit broken code” and “Fix Broken Builds Immediately’ present notable opportunities for monitoring CI practices; 2) developers do not adequately monitor all CI practices and express interest in monitoring additional practices; and 3) the most popular CI services currently offer limited native support for monitoring CI practices, requiring the use of third-party tools. Our results lead us to conclude that monitoring CI practices is often overlooked by both CI services and developers. Using third-party tools in conjunction with CI services is challenging, they monitor some redundant practices, and they still fall short of fully supporting CI practices monitoring. Therefore, CI services should implement CI practice monitoring, which would facilitate and encourage developers to monitor them.},
journal = {Empirical Software Engineering},
month = jun,
numpages = {47},
keywords = {Continuous integration, Monitoring, Software quality}
}

@inproceedings{elazharyICSME2019,
  author       = {Omar Elazhary and
                  Margaret{-}Anne D. Storey and
                  Neil A. Ernst and
                  Andy Zaidman},
  title        = {Do as {I} Do, Not as {I} Say: Do Contribution Guidelines Match the
                  {GitHub} Contribution Process?},
  booktitle    = {2019 {IEEE} International Conference on Software Maintenance and Evolution,
                  ({ICSME})},
  pages        = {286--290},
  publisher    = {{IEEE}},
  year         = {2019},
  -url          = {https://doi.org/10.1109/ICSME.2019.00043},
  -doi          = {10.1109/ICSME.2019.00043},
  timestamp    = {Tue, 05 Aug 2025 22:40:25 +0200},
  biburl       = {https://dblp.org/rec/conf/icsm/ElazharySEZ19.bib},
  bibsource    = {dblp computer science bibliography, https://dblp.org}
}

@inproceedings{delicheh2024mitigating,
  title={Mitigating security issues in {GitHub} {Actions}},
  author={Delicheh, Hassan Onsori and Mens, Tom},
  booktitle={Proceedings of the 2024 ACM/IEEE 4th International Workshop on Engineering and Cybersecurity of Critical Systems (EnCyCriS) and 2024 IEEE/ACM Second International Workshop on Software Vulnerability},
  pages={6--11},
  year={2024}
}

@inproceedings{koishybayev2022characterizing,
  title={Characterizing the security of {GitHub} {CI} workflows},
  author={Koishybayev, Igibek and Nahapetyan, Aleksandr and Zachariah, Raima and Muralee, Siddharth and Reaves, Bradley and Kapravelos, Alexandros and Machiry, Aravind},
  booktitle={31st USENIX Security Symposium (USENIX Security 22)},
  pages={2747--2763},
  year={2022}
}

@inproceedings{saroar2023developers,
  title={Developers’ perception of {GitHub} {Actions}: A survey analysis},
  author={Saroar, Sk Golam and Nayebi, Maleknaz},
  booktitle={Proceedings of the 27th International Conference on Evaluation and Assessment in Software Engineering},
  pages={121--130},
  year={2023},
  publisher={ACM}
}

@inproceedings{zhangICSE2024,
author = {Zhang, Yang and Wu, Yiwen and Chen, Tingting and Wang, Tao and Liu, Hui and Wang, Huaimin},
title = {How do Developers Talk about {GitHub} {Actions}? {Evidence} from Online Software Development Community},
year = {2024},
-isbn = {9798400702174},
publisher = {ACM},
-address = {New York, NY, USA},
-url = {https://doi.org/10.1145/3597503.3623327},
-doi = {10.1145/3597503.3623327},
abstract = {Continuous integration, deployment and delivery (CI/CD) have become cornerstones of DevOps practices. In recent years, GitHub Action (GHA) has rapidly replaced the traditional CI/CD tools on GitHub, providing efficiently automated workflows for developers. With the widespread use and influence of GHA, it is critical to understand the existing problems that GHA developers face in their practices as well as the potential solutions to these problems. Unfortunately, we currently have relatively little knowledge in this area. To fill this gap, we conduct a large-scale empirical study of 6,590 Stack Overflow (SO) questions and 315 GitHub issues. Our study leads to the first comprehensive taxonomy of problems related to GHA, covering 4 categories and 16 sub-categories. Then, we analyze the popularity and difficulty of problem categories and their correlations. Further, we summarize 56 solution strategies for different GHA problems. We also distill practical implications of our findings from the perspective of different audiences. We believe that our study contributes to the research of emerging GHA practices and guides the future support of tools and technologies.},
booktitle = {Proceedings of the IEEE/ACM 46th International Conference on Software Engineering (ICSE)},
-articleno = {42},
-numpages = {13},
keywords = {GitHub actions, empirical study, stack overflow},
-location = {Lisbon, Portugal},
-series = {ICSE '24}
}

@book{dekking2005modern,
	author = {Dekking, Frederik Michel and Kraaikamp, Cornelis and Lopuha{\"a}, Hendrik Paul and Meester, Ludolf Erwin},
	publisher = {Springer},
	title = {A Modern Introduction to Probability and Statistics: Understanding why and how},
	volume = {488},
	year = {2005}}

@book{cohen2013applied,
  title={Applied multiple regression/correlation analysis for the behavioral sciences},
  author={Cohen, Jacob and Cohen, Patricia and West, Stephen G and Aiken, Leona S},
  year={2013},
  publisher={Routledge}
}

@article{khatami2023state,
	author = {Ali Khatami and Andy Zaidman},
	title={State-of-the-practice in quality assurance in {Java}-based open source software development},
    author={Khatami, Ali and Zaidman, Andy},
    journal={Software: Practice and Experience},
    year={2024},
    publisher={Wiley},
    volume = {54},
    number = {8},
    pages = {1408--1446}
}

@inproceedings{gousios2014pullbased,
author = {Gousios, Georgios and Zaidman, Andy},
title = {A dataset for pull-based development research},
year = {2014},
-isbn = {9781450328630},
publisher = {ACM},
-address = {New York, NY, USA},
-url = {https://doi.org/10.1145/2597073.2597122},
-doi = {10.1145/2597073.2597122},
abstract = {Pull requests form a new method for collaborating in distributed software development. To study the pull request distributed development model, we constructed a dataset of almost 900 projects and 350,000 pull requests, including some of the largest users of pull requests on Github. In this paper, we describe how the project selection was done, we analyze the selected features and present a machine learning tool set for the R statistics environment.},
booktitle = {Proceedings of the 11th Working Conference on Mining Software Repositories},
pages = {368--371},
numpages = {4},
keywords = {pull-based development, pull request, empirical software engineering, distributed software development},
-location = {Hyderabad, India},
series = {MSR 2014}
}

@INPROCEEDINGS{bacchelli2013expectations,
  author={Bacchelli, Alberto and Bird, Christian},
  booktitle={2013 35th International Conference on Software Engineering (ICSE)}, 
  title={Expectations, outcomes, and challenges of modern code review}, 
  year={2013},
  volume={},
  number={},
  pages={712--721},
  keywords={Interviews;Inspection;Software;Context;Sorting;Guidelines;Knowledge transfer},
  -doi={10.1109/ICSE.2013.6606617}}

@inproceedings{khatamiSCAM2024b,
  author       = {Ali Khatami and
                  C{\'{e}}dric Willekens and
                  Andy Zaidman},
  title        = {Catching Smells in the Act: {A} GitHub Actions Workflow Investigation},
  booktitle    = {International Conference on Source Code Analysis and Manipulation (SCAM)},
  pages        = {47--58},
  publisher    = {{IEEE}},
  year         = {2024},
  -url          = {https://doi.org/10.1109/SCAM63643.2024.00015},
  -doi          = {10.1109/SCAM63643.2024.00015},
  timestamp    = {Wed, 15 Jan 2025 21:40:01 +0100},
  biburl       = {https://dblp.org/rec/conf/scam/KhatamiWZ24.bib},
  bibsource    = {dblp computer science bibliography, https://dblp.org}
}

\end{document}